\documentclass[aps,prl,preprint,superscriptaddress]{revtex4-2}

\usepackage{graphicx}

\usepackage{color}
\usepackage[normalem]{ulem}

\begin{document}

\title{Ferroelectric to paraelectric structural transition in LiTaO$_3$ and LiNbO$_3$}

\author{Felix Bernhardt}
\affiliation{Institut f\"ur Theoretische Physik, Justus-Liebig-Universit\"at Gie{\ss}en, Heinrich-Buff-Ring 16, 35392 Gie{\ss}en, Germany}
\affiliation{Center for Materials Research (ZfM/LaMa), Justus-Liebig-Universit\"at Gie{\ss}en, Heinrich-Buff-Ring 16, 35392 Gie{\ss}en, Germany}

\author{Leonard M. Verhoff}
\affiliation{Institut f\"ur Theoretische Physik, Justus-Liebig-Universit\"at Gie{\ss}en, Heinrich-Buff-Ring 16, 35392 Gie{\ss}en, Germany}

\author{Nils A. Sch\"afer}
\affiliation{Institut f\"ur Theoretische Physik, Justus-Liebig-Universit\"at Gie{\ss}en, Heinrich-Buff-Ring 16, 35392 Gie{\ss}en, Germany}

\author{Alexander Kapp}
\affiliation{Institut f\"ur Theoretische Physik, Justus-Liebig-Universit\"at Gie{\ss}en, Heinrich-Buff-Ring 16, 35392 Gie{\ss}en, Germany}

\author{Christa Fink}
\affiliation{Institut f\"ur Theoretische Physik, Justus-Liebig-Universit\"at Gie{\ss}en, Heinrich-Buff-Ring 16, 35392 Gie{\ss}en, Germany}
\affiliation{Center for Materials Research (ZfM/LaMa), Justus-Liebig-Universit\"at Gie{\ss}en, Heinrich-Buff-Ring 16, 35392 Gie{\ss}en, Germany}

\author{Wafaa \surname{Al Nachwati}}
\affiliation{Institut f\"ur Theoretische Physik, Justus-Liebig-Universit\"at Gie{\ss}en, Heinrich-Buff-Ring 16, 35392 Gie{\ss}en, Germany}

\author{Umar Bashir}
\affiliation{Leibniz-Institut f\"ur Kristallz\"uchtung, Max-Born-Str. 2, 12489 Berlin, Germany}

\author{Detlef Klimm}
\affiliation{Leibniz-Institut f\"ur Kristallz\"uchtung, Max-Born-Str. 2, 12489 Berlin, Germany}

\author{Fatima \surname{El Azzouzi}}
\affiliation{Institut f\"ur Energieforschung und Physikalische Technologien, Technische Universit\"at Clausthal, Am Stollen 19B, 38640 Goslar, Germany}

\author{{Uliana {Yakhnevych}}}
\affiliation{Institut f\"ur Energieforschung und Physikalische Technologien, Technische Universit\"at Clausthal, Am Stollen 19B, 38640 Goslar, Germany}

\author{{Yuriy {Suhak}}}
\affiliation{Institut f\"ur Energieforschung und Physikalische Technologien, Technische Universit\"at Clausthal, Am Stollen 19B, 38640 Goslar, Germany}

\author{Harald Schmidt}
\affiliation{Institut f\"ur Metallurgie, Technische Universit\"at Clausthal, 38678 Clausthal-Zellerfeld, Germany}

\author{Klaus-Dieter Becker}
\affiliation{Institut f\"ur Physikalische und Theoretische Chemie, Technische Universit\"at Braunschweig, 38106 Braunschweig, Germany}

\author{Steffen Ganschow}
\affiliation{Leibniz-Institut f\"ur Kristallz\"uchtung, Max-Born-Str. 2, 12489 Berlin, Germany}

\author{Holger Fritze}
\affiliation{Institut f\"ur Energieforschung und Physikalische Technologien, Technische Universit\"at Clausthal, Am Stollen 19B, 38640 Goslar, Germany}

\author{Simone Sanna}
\affiliation{Institut f\"ur Theoretische Physik, Justus-Liebig-Universit\"at Gie{\ss}en, Heinrich-Buff-Ring 16, 35392 Gie{\ss}en, Germany}
\affiliation{Center for Materials Research (ZfM/LaMa), Justus-Liebig-Universit\"at Gie{\ss}en, Heinrich-Buff-Ring 16, 35392 Gie{\ss}en, Germany}


\date{\today}

\begin{abstract}
The ferroelectric to paraelectric phase transition in LiTaO$_3$ and in pure as well as Mg doped LiNbO$_3$
is investigated theoretically by atomistic calculations in the framework of the density functional theory, 
as well as experimentally by calorimetry and electrical conductivity measurements.
First principles models within the stochastic self-consistent harmonic approximation (SSCHA) allow to 
consider anharmonic effects and thus to obtain a realistic estimate of the Curie temperature $T_C$ of 
both ferroelectrics. \textit{Ab initio} molecular dynamics (AIMD) calculations performed on large supercells 
confirm the Curie temperatures estimated with the SSCHA approach. Moreover, they also suggest that the 
structural phase transition is a continuous process beginning at temperatures well below $T_C$. 
According to AIMD, significant ionic displacements occurr already at temperatures 
of about 100\,K and 300\,K below $T_C$ in LiTaO$_3$ and LiNbO$_3$, respectively. 
To asses whether and how far the ionic displacements affect the materials properties, the AIMD 
results are 
compared with measurements of the electrical conductivity and of the heat capacity across the phase 
transition. Our first principles calculations 
moreover show that Mg ions, a frequently employed dopant, raise the Curie temperature in LiNbO$_3$. 
\end{abstract}


\maketitle


\section{\label{sec:introduction}Introduction}

Lithium niobate (LiNbO$_3$, LN) and lithium tantalate (LiTaO$_3$, LT) are two isomorph 
ferroelectrics, which are among the most widely used electro-optic materials \cite{Weis85}.
LN is characterized by unusually large pyroelectric, piezoelectric, electro-optic, and 
photo-elastic coefficients \cite{Weis85}. The magnitude of these coefficients is less 
pronounced in LT, which features, however, higher thermal stability. As an example, at 
high temperatures LT exhibits a much lower increase in electrical conductivity than LN, 
when the oxygen partial pressure is decreased \cite{UlianaDez23}.
Lithium niobate and tantalate are birefringent, have useful acoustic wave properties 
\cite{TCLee03} and a rather large acousto-optic figure-of-merit. 
The wealth of physical effects and, more important, their magnitude, render LN and LT
ideal candidates for acoustic and optical applications, exploiting both their bulk 
and surface properties \cite{Sanna17surf}.

Despite this wide range of applications, many aspects of the physics underlying 
the properties of LN and LT are not fully understood. This is particularly true for 
the phase transition between the paraelectric and the ferroelectric phases of the 
materials. This lack of knowledge is problematic. From an academic point of view,
phase transitions are a fascinating topic which deserves an accurate investigation.
Morever, the investigation of the ferroelectric to paraelectric transition is 
of technological relevance. Indeed, devices such as, e.g., sensors, are
often operated at high temperatures, even close to the transition temperature.


The ferroelectric to paraelectric structural transition in LiNbO$_3$ and LiTaO$_3$ is a phase 
transition of the second order \cite{Raeuber78,Volk08} according to the Ehrenfest notation. 
Accordingly, the spontaneous polarization $P_{S}$ steadily grows with decreasing temperature 
from 0 to a value of 71(62)\,$\mu$C/cm$^2$ and 60(55)\,$\mu$C/cm$^2$ 
for congruent (nearly stoichiometric) LiNbO$_3$ \cite{Chen01} and LiTaO$_3$ \cite{Kitamura98}, 
respectively. 
{The onset temperature} is known as transition or Curie 
temperature $T_C$, and has been measured to be in the range 1413\,K (1140\,$^\circ$C) -- 1475\,K (1202\,$^\circ$C) 
for LiNbO$_3$ \cite{Chen01,Nakamura08} and in the range 874\,K (601\,$^\circ$C)
-- 958\,K  (685\,$^\circ$C) for LiTaO$_3$ \cite{Kitamura98,Kim1969},
respectively. The large scattering of the measured values is in part due to the intrinsic difficulty
to perform measurements at about 1500\,K and in part to the nature of the phase transition, which 
is not a singular event occuring at $T_C$ but rather a continuous transition taking place in 
a temperature interval around $T_C$ \cite{6306010,Phillpot04}.

It has been controversially discussed whether the transition is of displacive or order-disorder type. 
A typical signature of displacive transitions is the presence of one or more optical phonon modes,
which become soft close to the Curie temperature $T_C$. No soft modes exist in order-disorder
transitions, instead. Therefore, many different studies have been focused on the investigation of the 
phonon modes of LN \cite{SimoRaman15,Samuel12,Friedrich15,Ridah97} and LT 
\cite{SimoRaman15,Friedrich16,Caciuc01,Rapitis88}. Some of the investigations, including 
Rayleigh scattering, Raman spectroscopy and infrared reflectivity, demonstrated the existence of a
$A_1$(TO) optical phonon becoming soft at high temperatures, suggesting a displacive nature of the
transition \cite{Barker67,Johnston68,Servoin79}. This interpretation was corroborated by Wood 
\textit{et al.}, who measured the birefringence of lithium niobate tantalate crystals 
(LiNb$_{1-x}$Ta$_x$O$_3$)  for various compositions and 
temperatures, finding that it changes continuously across the Curie temperature, as expected for a
displacive-type transition \cite{Wood08}. Yet, no mode softening could be observed in other 
studies, including neutron and Raman scattering experiments, suggesting the order-disorder nature of 
the ferroelectric to paraelectric phase transition \cite{Chowdhury74,Kojima99,Penna76}. Theoretical 
investigations based on the modeling of the phonon modes within the frozen phonon approach supported 
an order-disorder model for the oxygen atoms as the driving mechanism for the ferroelectric 
instability~\cite{Inbar96,Yu97}. 

This apparent contradiction was solved later by \textit{ab initio} \cite{6306010} and classic 
\cite{Phillpot04} molecular dynamics simulations, which showed that the ferroelectric to paraelectric 
phase transition is (at least in LiNbO$_3$) of both order-disorder 
and displacive type. More in detail, the calculations
demonstrated that the structural transformation is a process involving a displacive transition of 
the Nb sublattice and an order-disorder transition in the Li-O planes, 
which is completed at about $T_C$. Thus, the phase transition is a continuous process that takes place 
within a larger temperature interval and not an abrupt one as known, e.g., for solid to liquid
structural transitions.

Although the contribution of the atomistic simulations has been crucial for the interpretation of
the phase transition, the insight obtained in these studies must be considered of qualitative 
nature. \textit{Ab initio} molecular dynamics simulations have been performed in smaller 
supercells \cite{6306010}, thus failing to accurately estimate the transition temperature. 
Finite size effects indeed lead to the underestimation of $T_C$. Classic molecular dynamics 
can deal with larger supercells, however they have limited predictive power. Calculations
performed using different potentials predict a different behavior of the Nb-sublattice below 
the Curie temperature \cite{Phillpot04,Lee10}. Moreover, the theoretical investigations
performed so far neglected the thermal expansion, which might be a problematic approximation
for high values of $T_C$. For these reasons, the exact theoretical value of $T_C$ 
for LiNbO$_3$, LiTaO$_3$ and, more important, the temperature range in which the structural 
transformation occurs, are still to be determined.

An usual way to calculate the transition temperature from first principles is based on 
the estimate of the free energy of the two phases \cite{Simo_BA}, with the vibrational 
contribution to the entropy accounted for in harmonic approximation. This procedure has 
been attempted also for LiNbO$_3$ and LiTaO$_3$ as well \cite{Friedrich16}. However, the 
presence of imaginary phonon modes in the paraelectric structure does not allow the 
accurate calculation of the transition temperature. Moreover, the harmonic approximation
is questionable at temperatures close to $T_C$ of the investigated system. A recent 
study pointed out the
crucial role of anharmonicity in understanding the high temperature behavior, e.g., in 
thermal transport, of LT and, even more, of LN \cite{D1CE01323H}.

In this work we employ two different approaches to investigate the behavior of LiNbO$_3$ and LiTaO$_3$ 
in the temperature interval enclosing the Curie temperature. On the one hand, we employ the stochastic
self-consistent harmonic approach (SSCHA) for a realistic estimate of the transition temperature. 
On the other hand, we perform \textit{ab initio} molecular 
dynamics (AIMD) simulations on large supercells to estimate the temperature interval 
at which the structural transition occurs and investigate the mechanisms of the phase transition itself. 
We consider thereby the thermal expansion of the 
crystals. Both approaches can take anharmonicity into account and allow to investigate differences 
and similarities between LiNbO$_3$ and LiTaO$_3$. 

Calculations within the SSCHA approach yield a reliable phonon dispersion of the two phases and a 
realistic assessment of the transition temperature, which is calculated at 808\,K for LiTaO$_3$ and
at 1408\,K for LiNbO$_3$.
AIMD models confirm that the phase transition is a complex process involving 
a displacive transition of the Nb ions with the oxygen octahedra at a temperature below $T_C$ and an 
order-disorder transition in the Li ions, which is completed at $T_C$. In particular substantial ion 
displacement is predicted at temperatures below $T_C$ of about 100\,K for LiTaO$_3$ and 300\,K for 
LiNbO$_3$.
This temperature interval is further investigated by transport and
calorimetry experiments. The measurements reveal modifications of the 
conductivity in a temperature range similar to the calculated interval, while the heat capacity
continuously varies from the value of the ferroelectric phase to the value of the paraelectric phase 
in a more restricted interval.

It is moreover predicted that Mg doping, often used to enhance the resistance to optical damage, 
raises the transition temperature of stoichiometric LiNbO$_3$. Density functional theory calculations
within the nudged elastic band (NEB) method reveal that during the phase transition Mg has to overcome 
an energy barrier which is at least an order of magnitude larger than for the Li atoms, thus locally
pinning the ferroelectric phase. This confirms previous dielectric 
measurements of the Curie temperature as a function of the MgO content \cite{Grabmaier86}.

\section{\label{sec:methods}Methodology}

\subsection{\label{sec:sscha}Stochastic self-consistent harmonic approximation}
For the accurate estimate of the Curie temperature, 
we calculate the free energy $F$ per unit cell of the ferroelectric and 
of the paraelectric phase as a function of the 
temperature. The free energy per unit cell is given as:

\begin{equation}
\label{F}
    F=F_{el.}+F_{phon.} = U_{el.}+\sum_{\mathbf{q},\nu}\left[\frac{1}{2}\hbar\omega(\mathbf{q},\nu)+k_{\mathrm{B}} T \ln\left(1-\exp\left\{\frac{-\hbar\omega(\mathbf{q},\nu)}{k_{\mathrm{B}} T}\right\}\right)\right]
\end{equation}

where $\omega(\textbf{q}, \nu)$ are the phononic frequencies of mode $\nu$ at q-point 
$\textbf{q}$. We note that the frequencies $\omega(\mathbf{q},\nu)$ parametrically depend 
on the unit cell volume, which is
itself a function of temperature. Due to the large fundamental gap of LiTaO$_3$ and
LiNbO$_3$, the 
electronic entropy can be neglected in first approximation \cite{Riefer2013}. 

To estimate energies and frequencies in equation \ref{F}, the DFT as implemented in VASP 
\cite{Kresse1993,Kresse1996,Kresse1996_2} is employed. PAW potentials 
\cite{Bloechl94,Joubert1999} with PBEsol exchange-correlation functional \cite{Perdew2008} 
and electronic configurations 1s$^2$2s$^1$, 4p$^6$4d$^3$5s$^2$, 5s$^2$5p$^6$5d$^3$6s$^2$ and 2s$^2$2p$^4$ 
for Li, Nb, Ta and O respectively, are used. Our calculations include collinear spin-polarization. 
Hubbard corrections according to the simplified approach proposed by Dudarev
\textit{et al.} \cite{PhysRevB.57.1505} are applied to the $d$ orbitals of Ta and Nb.
Thereby we employ effective $U$ values of $U_{\mathrm{eff}}$ = 4\,eV for Nb and $U_{\mathrm{eff}}$ = 5\,eV for Ta,
according to reference \cite{Krampf_2021}.
Rhombohedral unit cells with $R3c$ and $R\overline{3}c$ symmetries model the ferroelectric
and paraelectric phase of LN and LT, respectively. A $6\times6\times6$ Monkhorst-Pack K-point 
mesh \cite{Pack1977}, as well as an energy cutoff of 500\,eV are needed to converge 
electronic energies to 10\,meV. A Gaussian smearing with width 0.02\,eV is applied to 
the Fermi occupancies. The ionic positions are optimized, such that all forces acting 
on the ions are lower than 0.005\,eV/{\AA}.

\begin{figure}[t]
  \includegraphics[width=0.3\textwidth,trim=0 0 225 0,clip]{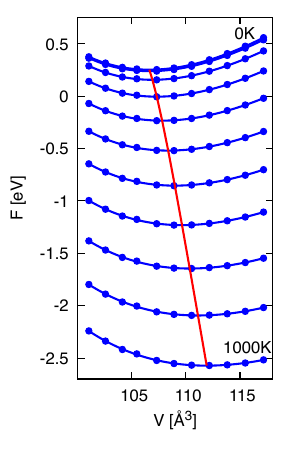}
  \includegraphics[width=0.3\textwidth,trim=0 0 225 0,clip]{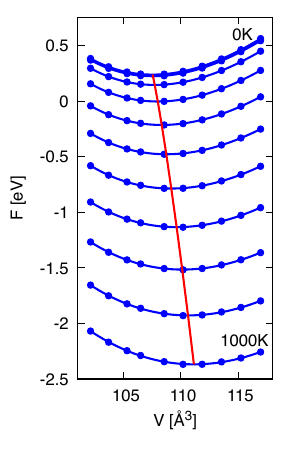}
  \caption{Gibbs free energy for different temperatures vs. unit cell volume of LiTaO$_3$. The 
    left side contains data from the ferroelectric phase, the right side for the 
    paraelectric phase. The datapoints are fitted to the Murnaghan-Birch equation
    of state, its minimum denotes the equilibrium volume at the corresponding  
    temperature. The red line serves as a guide to the eye.\label{QHA_fit}}
\end{figure}

In order to obtain the unit cell volume as a function of temperature, we employ the 
quasi-harmonic approximation (QHA)
as implemented in phonopy \cite{Togo2010, Togo2023}. Following the procedure 
outlined in Ref.~\cite{Togo2015}, we calculate harmonic phonon frequencies using 
the finite differences method with $3\times 3\times 3$ supercells at different 
volumina. Using the Parlinski-Li-Kawazoe method to interpolate the phonon 
frequencies to arbitrary q-points \cite{Parlinski1997}, we obtain the harmonic 
phonon frequencies on a $20\times 20\times 20$ mesh. This yields phononic free 
energies converged within 1\,meV with respect to a twice as dense q-point mesh. 
These frequencies and volumina are then used to calculate the Gibbs energy at 
different temperatures (assuming no internal pressure, see Eq. \ref{G}). Fitting the 
Gibbs energy via the Murnaghan-Birch equation of state, we obtain the equilibrium 
volume as a function of temperature. This is shown exemplarily for LiTaO$_3$ in 
figure \ref{QHA_fit}. The corresponding data for LiNbO$_3$, as well as a table with 
the unit cell volume and the lattice parameters calculated for each considered 
temperature is shown in the SI.

\begin{equation}
\label{G}
    G(T)=\min\limits_V\left[U_{el.}(V)+F_{phon.}(T;V)\right]
\end{equation}

Here, we consider rather high temperatures. It is therefore not 
reasonable to assume the validity of the harmonic approximation. Furthermore, the 
harmonic phonon frequency spectrum of the paraelectric (PE) phase of LT contains imaginary modes 
\cite{Friedrich2016}. An approach beyond the harmonic estimate of the phonon
frequencies, which includes thermal fluctuations, is needed. 

Thermal fluctuations (in particular anharmonic fluctuations) are often neglected in atomistic 
calculations. However, they affect the atomic structure, the phonon spectrum and thus the free 
energy of crystalline solids. The stochastic self-consistent harmonic approximation (SSCHA) method 
is an approach that allows to perform crystal geometry relaxation on the quantum mechanically 
calculated free energy landscape, optimizing the free energy with respect to all degrees of 
freedom of the crystal structure \cite{Monacelli2021}. In order to account for nuclear thermal fluctuations, 
however, the full Born-Oppenheimer energy surface, as well as its derivatives with respect to ionic 
positions (forces) and cell parameters (stress tensor) must be known, i.e., they must be calculated, e.g., 
by DFT within the stochastic approach as described in the following. Within this approach, we get 
access to the thermodynamics of crystals accounting for nuclear thermal anharmonic 
fluctuations. 

We use the SSCHA \cite{Errea2013,Errea2014,Bianco2017,Monacelli2018} as implemented 
in python-sscha \cite{Monacelli2021} to include anharmonic phononic effects. The 
formalism being used is described in detail 
in Ref.\cite{Monacelli2021}. We adopt the nomenclature established therein. 

In a first step, to sample the free energy landscape, a set of random ionic 
configurations is created in a chosen supercell according to the Gaussian 
probability distribution for the ions. Due to the severe computational cost, 
we only calculate anharmonic effects for selected temperatures using 
$2\times 2\times 2$ supercells (80 atoms). The equilibrium volume at a 
given temperature is assumed to be the one calculated within the QHA.

For PE LT and LN, we generate up to 400 structures per ensemble (see SI). As 
ferroelectric (FE) LT and LN have a 
lower symmetry, a larger ensemble size is necessary. Here, we use up to 
five times as many structures as for paraelectric LT or LN. On average, four ensembles 
are needed to find the free energy minimum. Lastly, another ensemble is created 
to calculate the anharmonic phonon frequencies. This last ensemble includes 
10000 structures, in order to achieve convergence of the phonon frequencies 
of around 5\,cm$^{-1}$, compared to a twice as large ensemble size. We assume 
the so called bubble-approximation (see reference \cite{Monacelli2021}), 
as its first order correction has only a 
minimal effect on the phonon frequencies. Again, these frequencies are then 
interpolated onto a $20\times 20\times 20$ mesh. The SSCHA calculated phonon
dispersion for LiTaO$_3$ and LiNbO$_3$ is shown in the SI.

The electronic energy is calculated using a stochastic average over this last 
ensemble, where the electronic energy of each structure is weighted according 
to the given temperature and dynamical matrix. The resulting stochastic uncertainty
of the electronic energies is less than 1\,meV.

Finally, inserting the calculated anharmonic phonon frequencies and electronic energies 
into Eq.\ref{F}, we can compare the free energy of both phases as a function 
of temperature.

\subsection{Molecular dynamics}
\label{sec:md}
While standard DFT calculations model atomic systems at 0\,K, molecular dynamics runs
allow to model atomic structures at finite temperatures by coupling the system with a
thermostat. The latter initializes the velocities of the atoms according to the 
Maxwell-Boltzmann distribution at the considered temperature. The time evolution of the 
system is obtained by solving the Newton equation of motion by means of a Verlet algorithm
within a given time step. At each time step, the forces acting on the ions are calculated
within quantum mechanics. In our calculations, we employ a time step of 2\,fs and model
the time evolution for about 5\,ps. The first 1\,ps of the MD trajectories, in which 
the system reaches thermal equilibrium is discarded and not considered for the data 
analysis. For the calculations of the MD trajectories, the Nos\'e-Hoover thermostat 
\cite{Nose84,Holian95} is employed, which models a canonical $NTV$ ensemble.

AIMD calculations are performed within the numerical approach described in the previous 
section, i.e., the DFT as implemented in VASP \cite{Kresse1993,Kresse1996,Kresse1996_2}, 
in combination with PAW potentials \cite{Bloechl94,Joubert1999} and the PBEsol 
\cite{Perdew2008} exchange-correlation potential is employed. The wave functions are expanded in a plane 
wave basis up to 400\,eV. In order to avoid self-correlation effects due to the periodic
boundary conditions, very large supercells consisting of a 4$\times$4$\times$4 repetition
of the rhombohedral unit cell (640 atoms) are employed. Accordingly, the volume of the 
Brillouin zone is rather limited and the energy integration is performed at the zone 
center ($\Gamma$-point). The volume of the cells in real space is 
calculated within the QHA at each considered temperature, as described in the previous 
section. Test calculations reveal that neglecting the thermal expansion and using
supercell of smaller size lead to results that are qualitatively similar.
However, a serious underestimation of the transition temperature occurs as 
discussed in the next section.  

From the AIMD runs a real time polarization can be extrapolated. Strictly speaking, the
macroscopic polarization $P_S$ for a supercell must be calculated within the modern theory
of polarization as a Berry phase of the Bloch orbitals \cite{Resta94,KS93,Resta93}. 
Yet, this approach is time consuming and cannot be applied to thousands of configurations 
of supercells containing 640 atoms. We employ therefore a simplified, approximated
approach, in which the macroscopic polarization is defined as dipole moment per volume 
unit calculated with respect to a reference phase with $P_S=0$ (e.g., the paraelectric phase).
More details about the calculation of $P_S$ are given in the SI. 

\subsection{\label{sec:berlin}Sample growth and calorimetry}

A LiTaO$_{3}$ crystal was grown via induction heating using the Czochralski process. 
The process was carried out in a protective argon atmosphere with a small addition of 
oxygen (less than 1 vol\,\%). The starting materials were mixtures of lithium carbonate 
(Li$_2$CO$_3$, Alfa Aesar, 5N) and tantalum pentoxide (Ta$_{2}$O$_{5}$, Fox Chemicals, 4N). 
The crystal was grown along the $c$-axis at a 0.5\,mm/h rate. Samples cut from the
grown material are labeled by IKZ in the following.
For comparison, a commercial Lithium Tantalate (LT) sample was utilized in this study. 
A wafer of congruent composition was purchased from Precision MicroOptics Inc. (PMO, USA) and 
was cut into wafers of dimensions 6$\times$5$\times$0.5\,mm$^{3}$, with both X-cut and 
Z-cut orientations. Samples cut from the commercial wafer are labeled by PMO in the following.
Lithium Niobate (LN) purchased from the same company and cut in the 
same state as LT was also employed in this research.

Differential scanning calorimetry (DSC) was applied to determine $T_c$. Thereby, 
a NETZSCH STA 449C ''F3'' thermal analyzer was used. Following the ASTM E1269 
standard, three consecutive measurements were done under identical conditions 
with empty crucibles (reference and sample), Al$_2$O$_3$ powder as standard with 
known $c_p(T)$ values, and the powdered sample. Four subsequent measurements were 
conducted up to 1463\,K in a flow mixture of 40\,ml/min Ar and 20\,ml/min O$_2$, 
with isothermal sections for equilibration. Subsequently, these heating segments' 
$c_p(T)$ functions from the last three measurements were averaged and used for 
further evaluation.

\subsection{\label{sec:goslar}Conductivity}

To measure the electrical conductivity of LiNbO$_{3}$ and LiTaO$_{3}$ single crystals, 
platinum electrodes are deposited on the samples through screen printing. The thickness 
of the electrodes is around 3\,{\textmu}m. Subsequently, the pieces are annealed at 
1000\,$^\circ$C for about an hour with a 2\,K/min heating rate.

An AC impedance spectroscopy is performed within a frequency range of 1\,Hz to 1\,MHz 
with an excitation AC voltage of 50\,mV to measure the conductivity. An 
impedance/gain-phase analyzer (Solartron 1260, Ametek Scientific Instruments, Hampshire, UK) 
is used for this purpose. The measurements are carried out in air, in a tube furnace that can heat 
up to 1580\,K at atmospheric pressure, starting from room temperature with a heating rate 
of 1\,K/min.  Precise temperature control is maintained using a Type S thermocouple to 
measure sample temperature and a Pt100 thermoresistor to compensate for cold-end 
temperature variations of the thermocouple. 

Modeling the electrical properties of the 
samples was done using a single equivalent circuit consisting of a bulk resistor ($R_B$) 
in parallel with a constant phase element (CPE) based on the obtained single semicircular 
EIS features. The bulk conductivity  $\sigma$ is calculated using the formula 
$\sigma = t (A R_B)^{-1}$, where $t$ and $A$ represent the sample thickness and electrode 
area, respectively.

The activation energy $E_A$ is determined using the Arrhenius relation as follows:

\begin{equation}
	\label{eq:cond}
	\sigma=\sigma_0/T \exp \left(-E_A/(k_BT)\right)
\end{equation}

Here, $\sigma_0$, $k_B$ and $T$ represent a constant pre-exponential coefficient, 
the Boltzmann constant, and the absolute temperature, respectively. The Nernst-Einstein 
relation states that $1/T$ reflects the connection between mobility, which affects 
electrical conductivity, and thermally activated diffusion. Equation \ref{eq:cond} 
provides the activation energy $E_A$ if a single conduction mechanism dominates over an 
extensive temperature range, reflected by a constant slope in the Arrhenius representation. 
The logarithmic slope of $\sigma{T}$ in Eq. \ref{eq:slope} is used for analysis as the 
Arrhenius plot does not show differences in electrical conductivity at high temperatures 
(more details on the procedure can be fond, e.g., in Ref. \cite{Krampf_2021}).

\begin{equation}
	\label{eq:slope}
	E_{\sigma T} = -k_B \frac{\partial \ln{(\sigma T)}}{\partial (1/T)}.
\end{equation}

The consideration of $\sigma T$ is discussed in the following. 

\begin{figure}[t]
  \includegraphics[width=0.3\textwidth,trim=0 0 183 0,clip]{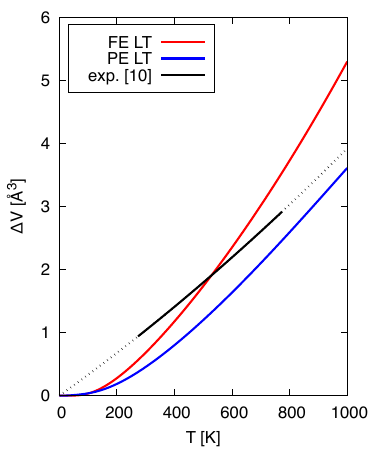}
  \includegraphics[width=0.3\textwidth,trim=0 0 183 0,clip]{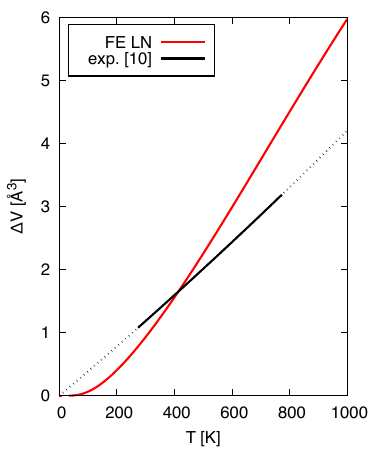}
  \caption{Unit cell volume expansion as a function of temperature for LiTaO$_3$ (lhs)
    and LiNbO$_3$ (rhs). The red and blue lines denote the thermal expansion of
    the ferro- and paraelectric configuration as calculated 
    within the QHA, respectively. The black line is a measurement performed
    in reference \cite{Kim1969}. \label{fig_vol_fit}} 
\end{figure}


\section{\label{sec:results}Results}

\subsection{\label{sec:res_sscha}Transition temperature}

The Curie temperature of the structural transition between the ferroelectric and 
the paraelectric phase is obtained inserting the phonon frequencies including 
anharmonic contributions as estimated in the SSCHA framework and the electronic 
energies into equation \ref{F}. The data are calculated considering the thermal expansion
as estimated within the QHA, as shown in figure \ref{fig_vol_fit}.
The calculated data are in overall good agreement with the measured thermal expansion
of LiTaO$_3$ and LiNbO$_3$ as given by the coefficients from Ref. \cite{Kim1969}.
The calculated values slightly overestimate the measured volumes in both materials 
by around 2\%,
as known from GGA based exchange and correlation functionals. The deviations might 
also be related to the stoichiometric composition of the measured samples or
by anharmonic contributions disregarded in the QHA.

\begin{figure}[t]
\centering
(a)\hspace{5cm}(b)\hspace{4cm}(c)\\
    \includegraphics[width=0.3\textwidth,trim=0 0 183 0,clip]{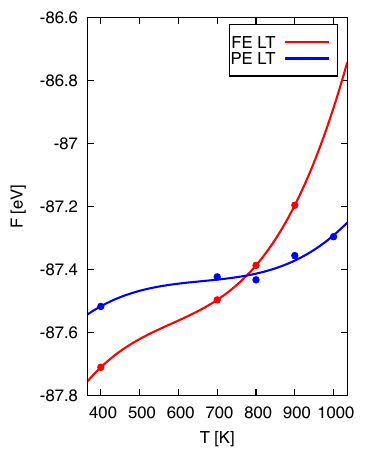}
    \includegraphics[width=0.3\textwidth,trim=0 0 183 0,clip]{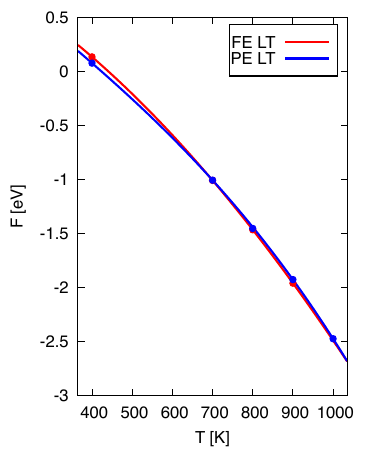}
    \includegraphics[width=0.3\textwidth,trim=0 0 183 0,clip]{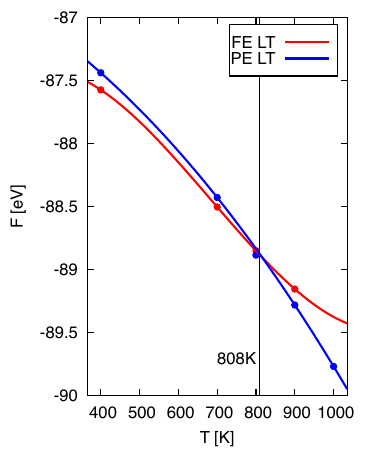}

    \includegraphics[width=0.3\textwidth,trim=0 0 183 0,clip]{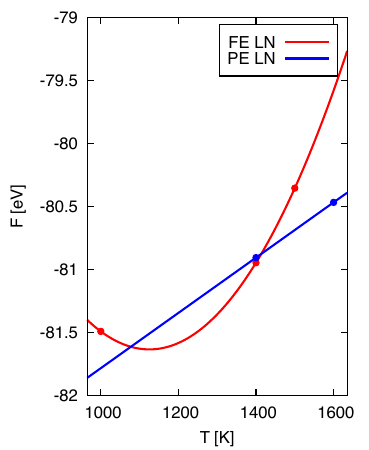}
    \includegraphics[width=0.3\textwidth,trim=0 0 183 0,clip]{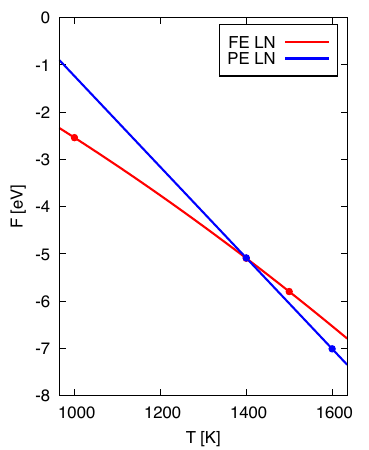}
    \includegraphics[width=0.3\textwidth,trim=0 0 183 0,clip]{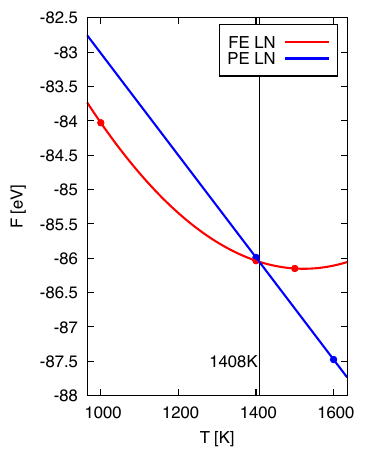}

    \caption{Free energy per unit cell calculated as a function of the temperature  
      for LiTaO$_3$ (upper row) and LiNbO$_3$ (lower row). The red and blue lines denote 
      the free energy of the FE- and PE configuration as calculated within the SSCHA, 
      respectively. (a) Electronic contribution, (b) vibrational contribution and 
      (c) total free energy are shown.\label{fig:free_en}}
\end{figure}

Figure \ref{fig:free_en} shows the free energy of the unit cell of the ferroelectric 
and paraelectric structure of LT (upper part) and LN (lower part) as a function 
of the temperature. Electronic and 
vibrational contributions (as calculated with equation \ref{F}) are shown in 
\ref{fig:free_en} (a) and \ref{fig:free_en} (b), respectively, while \ref{fig:free_en} 
(c) shows the total free energy. Assuming a minor dependence on the pressure (a resonable
approximation for solids), the groundstate configuration for a certain volume is given at each 
temperature by the structure with the lower free energy. The Curie temperature is  
the intersection of both curves. In our calculation the intersection occurs at 
around 808\,K and 1408\,K for LiTaO$_3$ and LiNbO$_3$, respectively. They are 
rather close to the experimentally determined value of 874\,K \cite{Kitamura98,Kim1969}
and 1413\,K  \cite{Chen01}, respectively, although the value calculated for LT
underestimates somehow the measured value.

The transition temperature of LiNbO$_3$ was estimated in Ref. \cite{Friedrich2016} 
comparing the free energies of the paraelectric and ferroelectric phases, similarly to the approach we apply, however within 
the harmonic approximation. The calculated transition temperature was estimated to be 1000\,K using 
the DFT calculated equilibrium volume and 1160\,K using the high-temperature experimental volume.
Our calculations based on the SSCHA formalism predict for LiNbO$_3$ a transition temperature of 
1408\,K and thus confirm not only the obvious consideration that quasi-harmonic contribution are 
of fundamental importance, but also allow to quantify their impact on the overall results. Comparing 
the results obtained neglecting and including non-harmonic effects, we can conclude that they cause 
a shift of the transition temperature of ca. 250\,K, greatly improving the agreement between measured 
and calculated transition temperatures.

\subsection{\label{sec:res_md}Transition dynamics}

After determining the Curie temperatures for the ferroelectric to paraelectric transition 
in LiTaO$_3$ and LiNbO$_3$ within the SSCHA approach, we adopt a microscopic perspective to
investigate the atomic displacements at the phase transition. This yields information
about the mechanisms of the phase transition at the atomic scale and about the width of the 
temperature interval at which the structural transition occurs.

\begin{figure}[t]
  \includegraphics[width=0.6\linewidth]{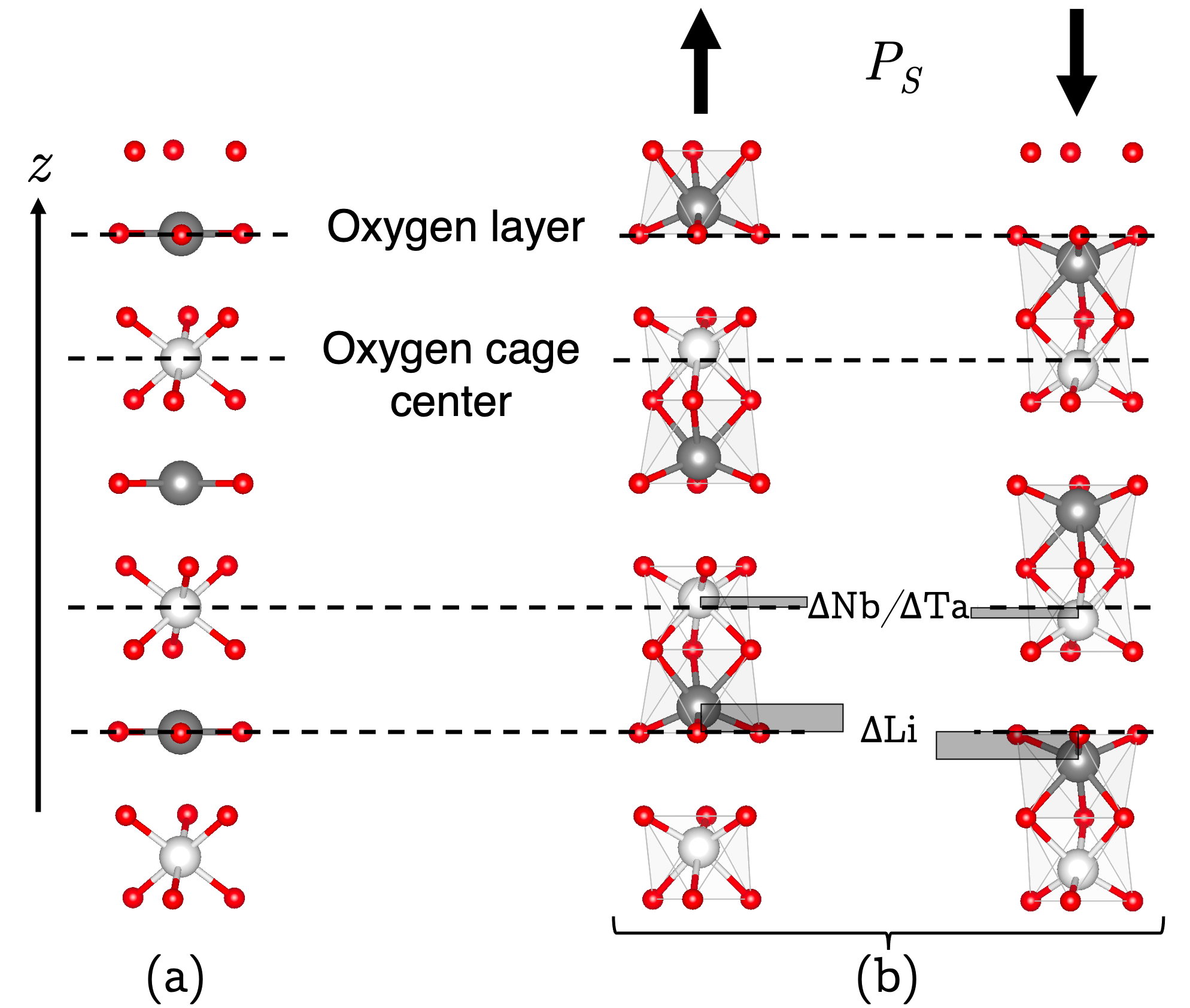}
  \caption{Atomic structure of the (a) paraelectric and (b) ferroelectric phase of LiNbO$_3$
    and LiTaO$_3$. Nb/Ta atoms are white, Li atoms gray and O atoms red. The displacement from
    the atomic positions in the paraelectric phase is indicated with $\Delta$Nb or $\Delta$Ta and $\Delta$Li, 
    respectively. Both displacements occur in the crystallographic $z$ direction, parallel to the spontaneous 
    polarization $P_S$. \label{fig:crystals}}
\end{figure}

In order to understand the atomic displacement at the structural transformation, it is crucial 
to describe the microscopic structure of the two phases involved in the transition. 
LiTaO$_3$ and LiNbO$_3$ can be both thought of as oxygen octahedra piled along the crystal 
$c$ axis. These octahedra may be empty or host Li and Ta/Nb atoms. In the ferroelectric 
structure, the octahedra occupation follows the order (from bottom to
top) Li, Ta/Nb, empty, Li, Ta/Nb, empty, and so forth, as shown in figure \ref{fig:crystals}.
The Nb/Ta atoms are not at the center of the oxygen octahedra, while the Li atoms are just
above (or below) an oxygen plane. This atomic arrangement is non-centrosymmetric (space
group $R3c$). The center of mass of the positive (Nb$^{5+}$, Ta$^{5+}$, Li$^+$) 
and negative (O$^{2-}$) charges are displaced, giving rise to a spontaneous polarization 
along the $c$-axis as large as 0.60 C/m$^2$ for LiTaO$_3$ \cite{Kitamura98} and 0.71 C/m$^2$ for 
LiNbO$_3$ \cite{Chen01}, as discussed in the introduction.

The high temperature phase of LiTaO$_3$ and LiNbO$_3$ (above 874\,K \cite{Kitamura98} and 
1411\,K \cite{Chen01}, respectively) 
is paraelectric and belongs to the space group $R\overline{3}c$. According to the general 
understanding of this structure, the Li ions lie exactly within the 
oxygen planes, while the Ta and Nb atoms sit exactly at the center of oxygen octahedra, as 
shown in figure \ref{fig:crystals}. The structure has a higher symmetry than the ferroelectric
phase. In particular, the centers of mass of the positive and negative charges coincide, and the
crystal shows no spontaneous polarization. We remark that  this picture has been previously 
questioned \cite{6306010}. It has been suggested that it only holds in time average, while the instantaneous
configuration features Li ions randomly distributed above or below the oxygen planes, so that 
oppositely directed microscopic dipole moments sum up to a zero net polarization \cite{6306010}.

The deviation of the atomic positions in the 
ferroelectric and paraelectric structure can be quantified by the displacement of the Nb/Ta 
ions from the oxygen cage center ($\Delta$Nb, $\Delta$Ta) and of the Li ions from the 
oxygen planes ($\Delta$Li). According to our calculations, at 0\,K (in the ferroelectric phase) 
the displacements assume the value $\Delta$Nb = 0.279 {\AA}, $\Delta$Li = 0.717 {\AA} in 
LiNbO$_3$, and  $\Delta$Ta = 0.188 {\AA}, $\Delta$Li = 0.643 {\AA} in LiTaO$_3$, respectively.

\begin{figure}[t]
\includegraphics[width=0.49\linewidth]{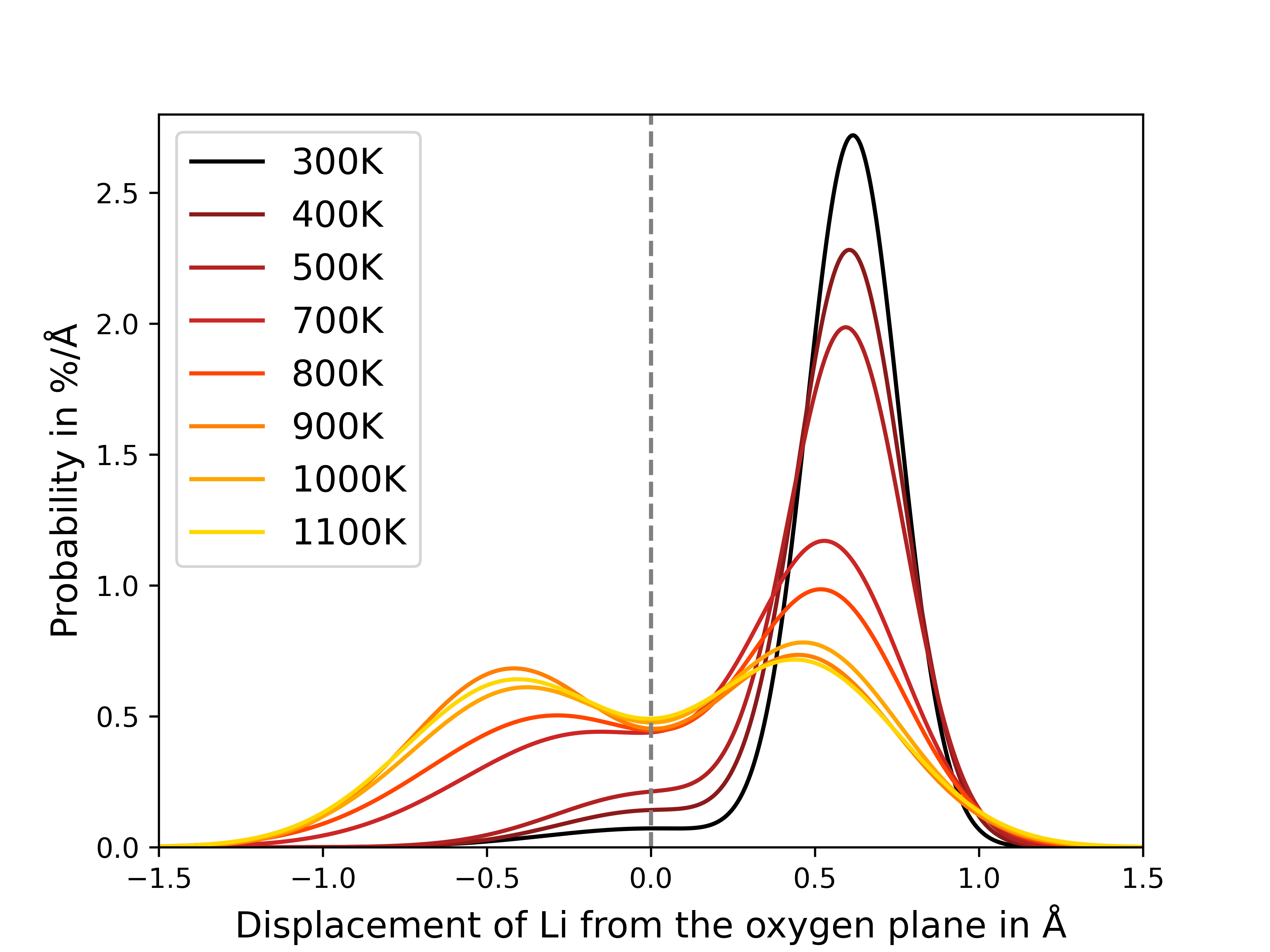}
\includegraphics[width=0.49\linewidth]{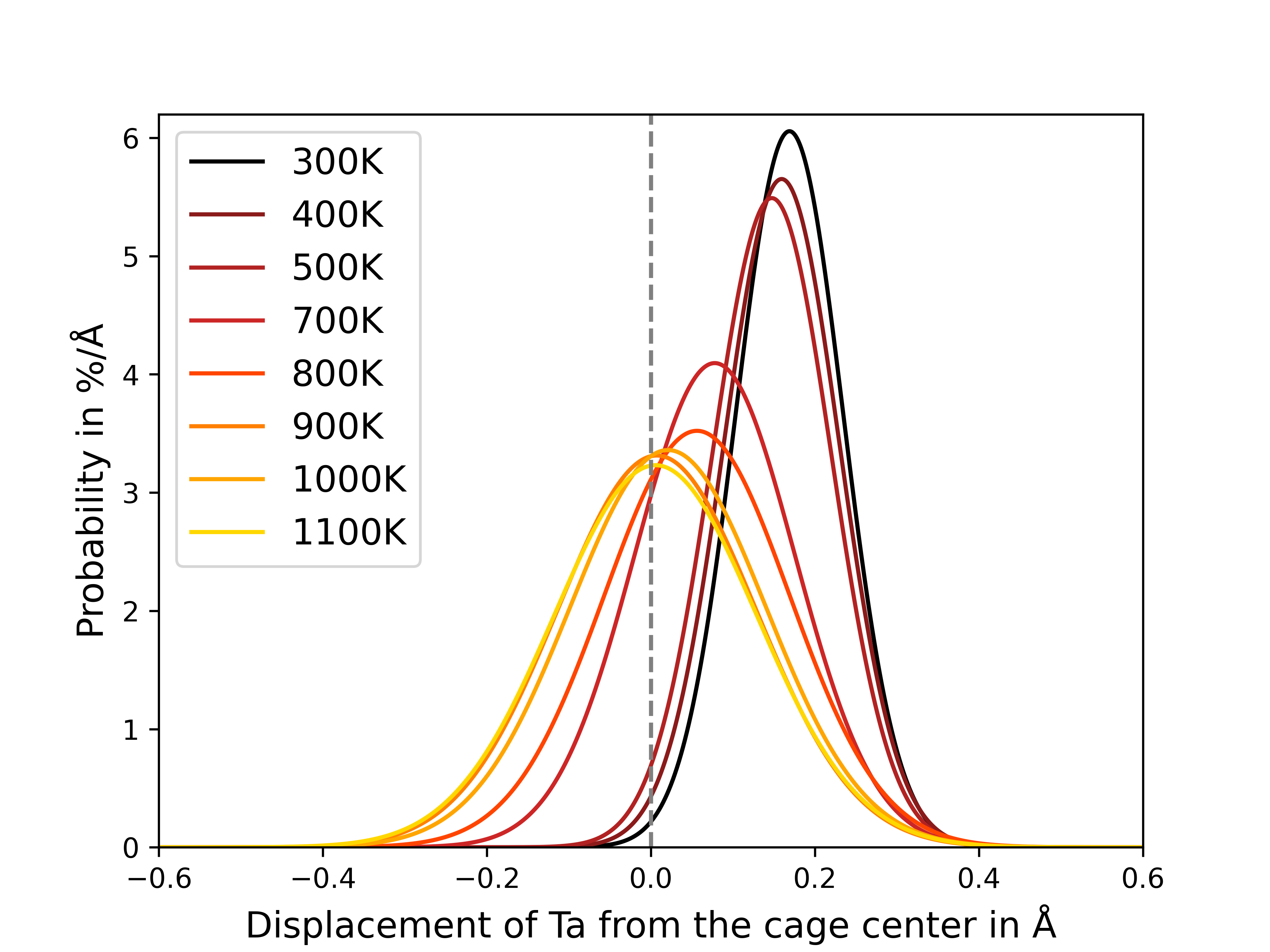}
\caption{(a) Temperature dependent probability distribution of the parameter 
$\Delta$Li (representing the displacement of the Li ions from the oxygen planes) 
for LiTaO$_3$.  The unimodal distribution well below the Curie temperature
(e.g., at 300\,K) means that all the Li ions are located above the oxygen planes.
The symmetric, bimodal curve at above the Curie temperature (e.g., at 1100\,K) 
indicates that the Li ions are randomly distributed above or under the oxygen planes, 
as expected for an order-disorder phase transition.
(b) Probability distribution of the parameter $\Delta$Ta (representing the 
displacement of the Ta ions from the center of the oxygen octahedra). The distribution
is unimodal at every temperature, as expected for a displacive phase transition.
\label{fig:displ_LT}}
\end{figure}

To investigate the dynamics of the phase transition, we estimate the temperature dependence
of the parameters $\Delta$Li and $\Delta$Nb, $\Delta$Ta, which we extract from the
AIMD runs performed at different temperatures. Let us start our discussion with LiTaO$_3$.
From the crystal structure shown in figure \ref{fig:crystals}, it can be expected that above 
the Curie temperature $\Delta$Li and $\Delta$Ta are distributed about a value of zero, 
while at low temperatures they are distributed around the value calculated for 0\,K of 
0.188\,{\AA} for $\Delta$Ta and 0.643\,{\AA} for $\Delta$Li, respectively.

In figure \ref{fig:displ_LT} the distributions of $\Delta$Li (lhs) and $\Delta$Ta (rhs) 
extracted from AIMD runs of LiTaO$_3$ at different temperatures are shown
(gaussian fit, the raw data are shown in the SI). The distribution
$\Delta$Li and $\Delta$Ta well below $T_C$ (e.g., at 300\,K, black curve) are indeed 
centered at the expected values. The form of the distribution is gaussian, as expected 
for the thermal broadening of the sharp peak of the ideal, rigid ferroelectric structure.
In particular, we observe that the $\Delta$Li
probability distribution is unimodal, suggesting that all the Li ions are above the 
oxygen planes, as expected for the ferroelectric phase. 

Figure \ref{fig:displ_LT} (rhs) shows the probability distribution of $\Delta$Ta at 
different temperatures (Gaussian fit). The distribution is unimodal at each 
investigated temperature, as expected for a displacive transition. With increasing 
temperature, the distribution broadens, since the atoms vibrate with larger amplitude
about the equilibrium positions. Moreover, the distribution smoothly moves toward 
$\Delta$Ta = 0. This can be 
interpreted as a consequence of the fact that Ta atoms can move relatively free within
the oxygen octahedra, as they do not need to overcome any energy barrier to move from 
the ferroelectric to the paraelectric configuration.

The temperature dependence of the probability distribution of $\Delta$Li, 
shown in figure \ref{fig:displ_LT} (lhs) is in 
striking contrast with the probability distribution of $\Delta$Ta. With growing 
temperatures, the distribution does not substantially shift toward $\Delta$Li = 0. 
Instead, it becomes bimodal. 
Well above the structural transition, the distribution is symmetric with respect to 
$\Delta$Li = 0 (oxygen lattice plane), which represents a local minimum. 
Only a small fraction of the Li atoms lies exactly in the plane,
independently from the temperature. While the oxygen plane has been regarded as an 
energetic barrier (at least for LiNbO$_3$) in the past \cite{Phillpot04}, it can also be 
stated that when the Li atoms oscillate around $\Delta$Li = 0 their velocity is maximal 
and their probability to be found at this point is minimal.

The calculated curves shown in figure \ref{fig:displ_LT} (lhs)
demonstrate that for temperatures well below $T_C$ almost all Li atoms are distributed about the
positions they assume in the ferroelectric structure. With increasing temperature a growing 
fraction of Li atoms possesses enough thermal energy to pass the oxygen plane. Above $T_C$
all Li atoms have sufficient energy to move across the oxygen plane, so that they are distributed
roughly with the same probability above or below an oxygen plane. The sum of all Li contributions
to the spontaneous polarization vanishes. However, $P_S$ is exactly zero only in average, 
as each unit cell still has a finite, randomly distributed, dipole moment. This feature is 
compatible with an order-disorder type structural transition. This is in 
agrement with a recent X-ray diffraction investigation shwowing that ordering of the disordered 
Li ion in the polar direction accompanied by deformation of the oxygen octahedra leads 
to the ferroelectric phase transition \cite{Zhang_2018}.

We observe that already at temperatures significantly lower than $T_C$ the probability
distributions for both $\Delta$Li and $\Delta$Ta (e.g., already at 700\,K) substantially 
differ from the distributions calculated for low temperatures (e.g., 300\,K). This suggests 
that the phase transition does not occur abruptly at a given temperature, but is rather a 
continuous process, occurring dynamically over a certain range of temperatures. The width
of this temperature range will be explored in the next sections.

\begin{figure}[t]
\includegraphics[width=0.46\linewidth]{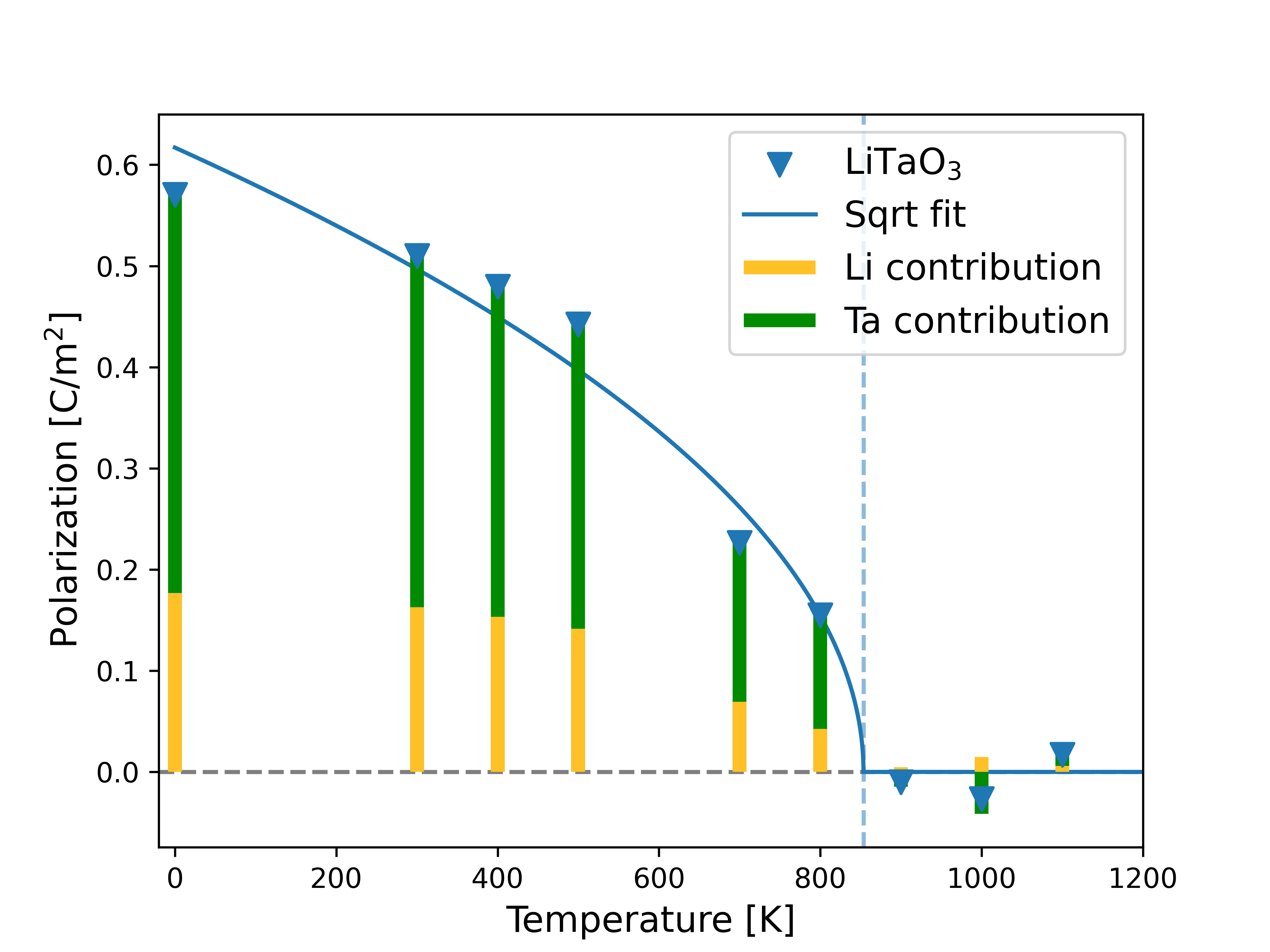}
\includegraphics[width=0.46\linewidth]{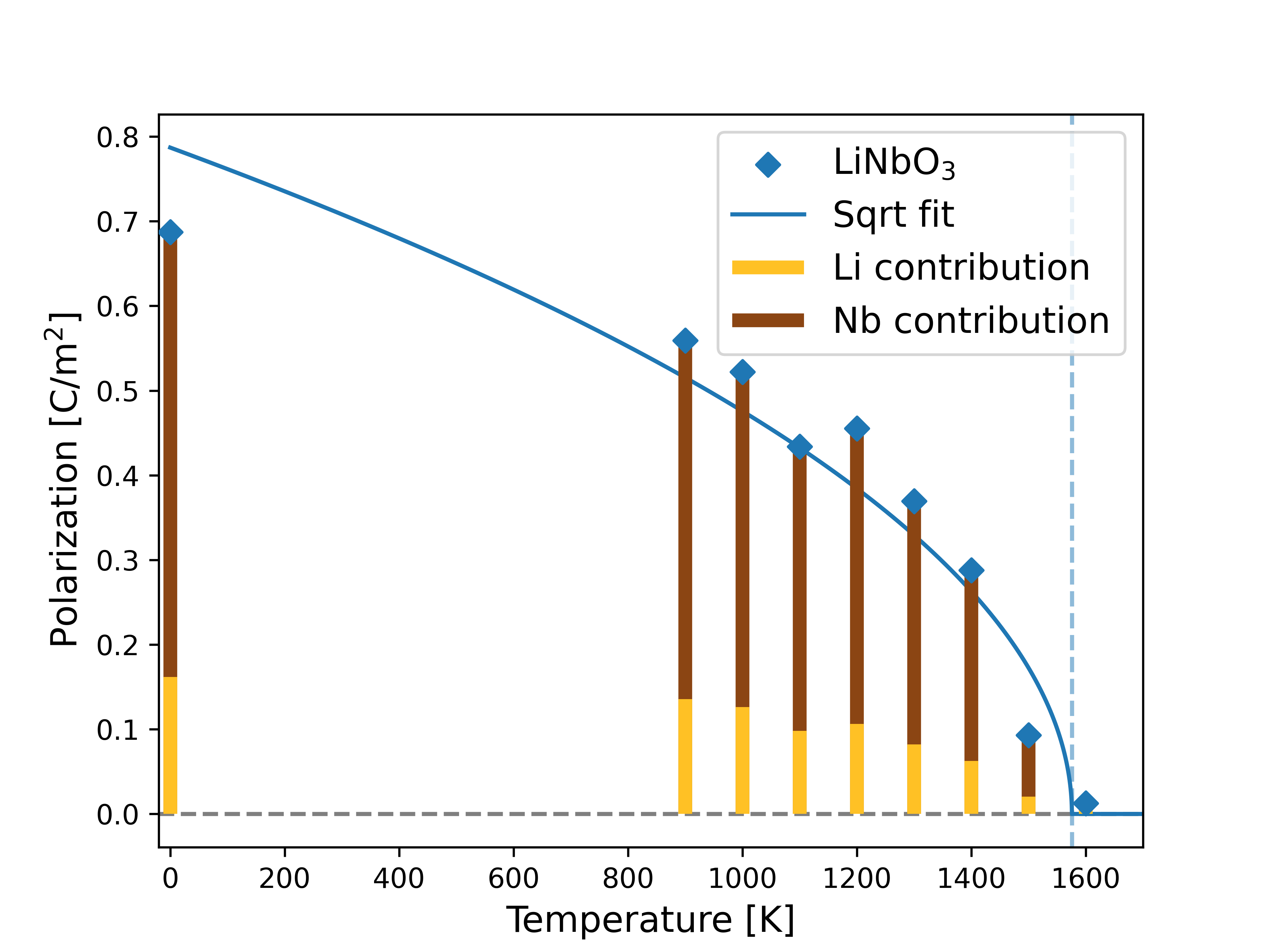}
\caption{Calculated temperature dependence of the spontaneous polarization $P_S$ for (a) LiTaO$_3$
and (b) LiNbO$_3$. The solid line is a square root fit of the calculated data, represented by the 
histograms. The latter are color coded according to the contribution of the Li sublattice
(in yellow) and of the Ta/Nb sublattice (green/brown). \label{fig:polarisation}}
\end{figure}

The instantaneous spontaneous polarization can be extrapolated from the AIMD snapshots 
with the simplified approach explained in the methodological section. The average polarization
of LiTaO$_3$ as a function of the temperature is shown in figure \ref{fig:polarisation}(a).
The calculated data can be fitted by a root function. The critical exponent of the order
parameter in a mean field theory such as the Landau-Ginzburg theory is 0.5, however lower 
exponents are calculated within other approaches. In this work, we chose a square root fit 
of the spontaneous polarization as a function of the temperature, as expected for a 
second-order phase transition in the Landau-Ginzburg theory. 
The square root function equals zero at 841\,K. At this temperature, 
the polarization vanishes and the crystal is in the paraelectric configuration.
This value is in excellent agreement with the experimentally determined transition 
temperature of 873-878\,K \cite{FatimaDez23,Kitamura98,Bashir23}.

The contribution to the polarization of the different species is also shown in 
figure \ref{fig:polarisation}(a). It can be observed that both the Li and Ta 
contribution are roughly a square root function of the temperature. The Ta atoms, which
experience limited shifts but carry a high (nominal) charge of $5+$ electrons, have a larger 
contribution to the total polarization than the Li atoms, which undergo larger 
displacements but carry a much lower (nominal) charge of $1+$ electrons.

\begin{figure}[t]
\includegraphics[width=0.49\linewidth]{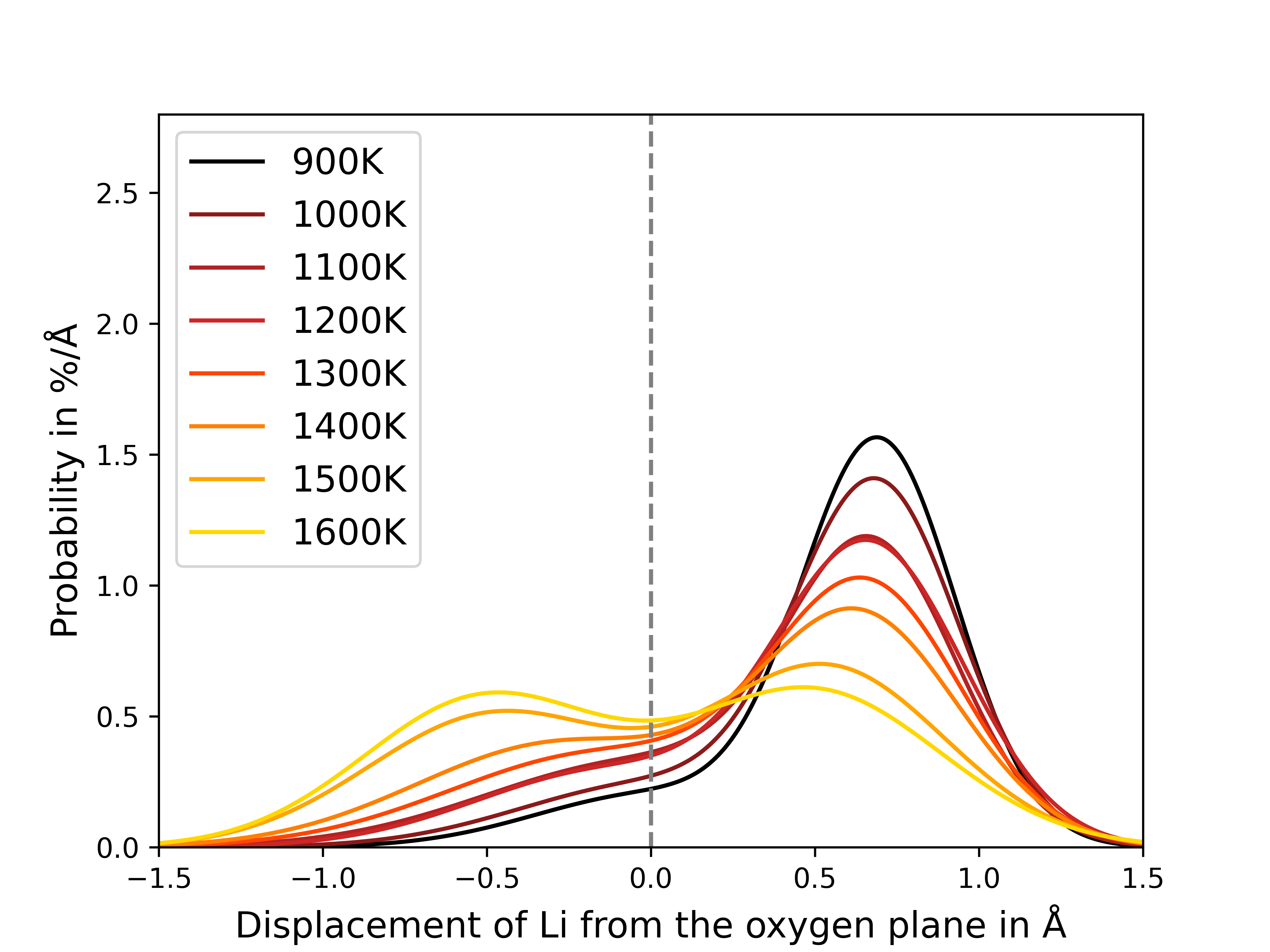}
\includegraphics[width=0.49\linewidth]{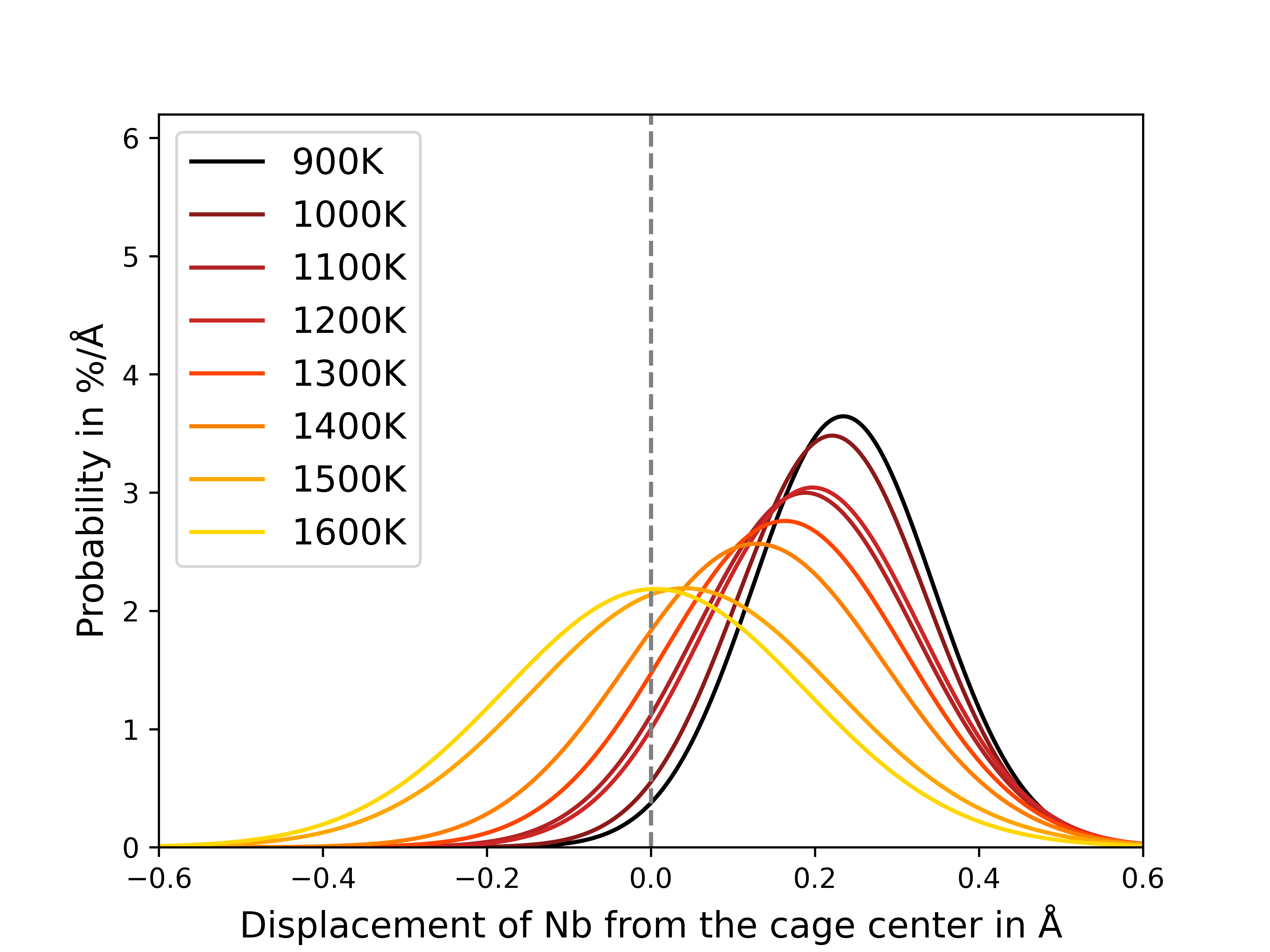}
\caption{(a) Temperature dependent probability distribution of the parameter 
$\Delta$Li (representing the displacement of the Li ions from the oxygen planes) 
for LiNbO$_3$.  The unimodal distribution well below the Curie temperature
(e.g., at 900\,K) means that all the Li ions are located above the oxygen planes.
The symmetric, bimodal curve at above the Curie temperature (e.g., at 1600\,K) 
indicates that the Li ions are randomly distributed above or under the oxygen planes, 
as expected for an order-disorder phase transition.
(b) Probability distribution of the parameter $\Delta$Nb (representing the 
displacement of the Nb ions from the center of the oxygen octahedra). The distribution
is unimodal at every temperature, as expected for a displacive phase transition.
\label{fig:displ_LN}}
\end{figure}

Summarizing, the AIMD calculations reveal that the ferroelectric to paraelectric 
transition in LiTaO$_3$ is a dynamical process of displacive type in the Ta 
sublattice and of order-disorder type in the Li sublattice. The square root fit of the 
spontaneous polarization vanishes at 841\,K, which can be considered as the 
$T_C$ value predicted by AIMD. A similar behavior 
has been previously proposed for LiNbO$_3$ \cite{6306010,Phillpot04}.

Figure \ref{fig:displ_LN} shows the temperature dependent probability distribution for
$\Delta$Li and $\Delta$Nb for LiNbO$_3$ extrapolated from the AIMD trajectories. The
overall behavior is qualitatively similar to LiTaO$_3$. The distributions can be fitted 
by one (Nb) or two (Li) gaussian curves, with the distribution $\Delta$Nb unimodal for all 
investigated temperatures. The distribution of $\Delta$Li is bimodal and again symmetric
with respect to the oxygen plane in the paraelectric configuration, and unimodal at low
temperatures. As the relevant temperatures in LiNbO$_3$ are higher than in LiTaO$_3$,
the gaussian distributions are broader. 
We point out that caution is due in the interpretation of the displacement
distribution at 1600\,K in figure \ref{fig:displ_LN}, as this temperature is beyond 
the melting point of LiNbO$_3$. Due to known deficiencies of MD in the estimation of 
the nucleation free energy barrier between the solid and the liquid phases \cite{ZOU2020109156},
the material is modelled at this temperature as an overheated phase.

The spontaneous polarization, shown in figure \ref{fig:polarisation}(b), is 
fitted by a square root function which vanishes at 1524\,K, which slightly 
overestimates the experimentally determined transition temperature of 1430-1475\,K 
\cite{FatimaDez23,Chen01,Bashir23}. Again, the Li sublattice has a smaller contribution 
to the spontaneous polarization than the Nb sublattice. Considering 
the thermal expansion is fundamental for a realistic estimate of the transition 
temperature. A comparison of the values calculated with and without thermal expansion
is given in the SI.

Our calculations for LiNbO$_3$ confirm the suggestion first made in \cite{6306010,Phillpot04} 
that the phase transition is of mixed displacive and order-disorder type, as it involves a
displacive transition of the Nb ions within the oxygen octahedra and an order-disorder 
transition of the Li sublattice. The atomic structure and the spontaneous polarization change 
gradually between the ferroelectric and paraelectric phase, so that the transition does not 
occur abruptly at a well defined temperature but rather over a temperature range of several 
100\,K, which will be further investigated later. 


\begin{figure}
  \includegraphics[width=0.7\textwidth]{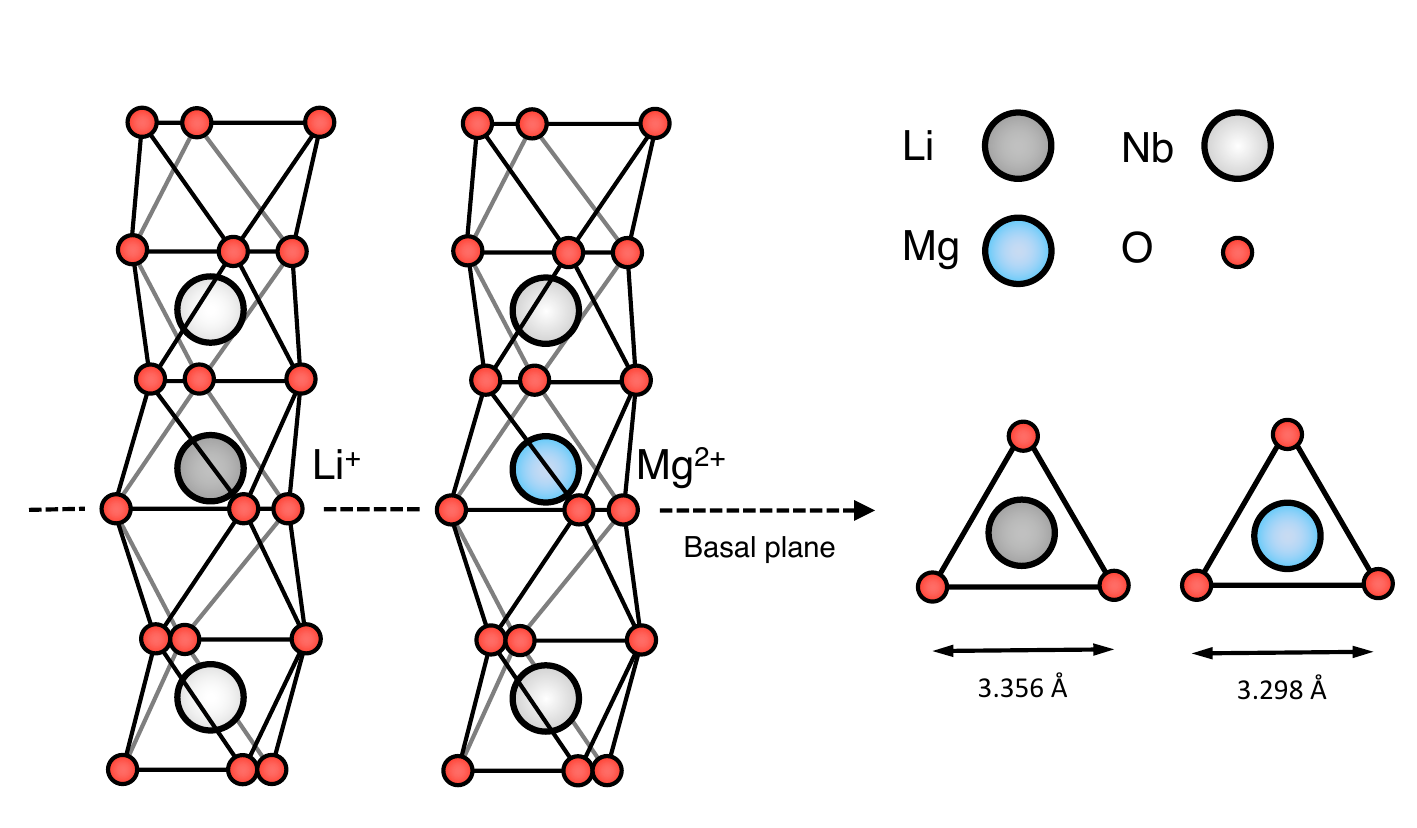}
  \caption{Schematic representation of the atomic structure of undoped and Mg doped
    LiNbO$_3$. The oxygen-oxygen interatomic distances are given.}\label{fig:mg_geom}
\end{figure}

\subsection{\label{sec:res_doping}Effect of Mg doping on the transition temperature} 

In a final step, we investigated how Mg doping affects the Curie temperature of LiNbO$_3$.
Mg is the chief dopant for LiNbO$_3$, if the optical damage resistance has to be 
enhanced. Indeed, a concentration of Mg exceeding a so called threshold of 
about 5\,mol\% leads to a reduction of the optical damage of more than two orders 
of magnitude \cite{Volk08}.
The mechanism leading to the optical resistance is quite indirect.
The Mg incorporation leads to the formation of Mg$^{2+}_{\text{Li}}$ substitutionals \cite{YanluMg},
which inhibit the formation of Nb$_{\text{Li}}$ antisites (small bound polarons) that 
are crucial for the photorefractivity. 
In LiTaO$_3$, doping with Mg lowers the photorefraction as well \cite{Nitanda95}, however, we limit 
our investigation to LiNbO$_3$.

Whether Mg increases or decreases $T_C$ depends on the sample composition (nearly 
stoichiometric or congruent) and on the Mg concentration \cite{Bryan84}. In our models, 
we consider a number of Mg$^{2+}_{\text{Li}}$ substitutionals in stoichiometric LN 
which corresponds to 
5.56\,mol.\%. This is a typical doping concentration of the LN samples used, e.g., 
for the fabrication of optical waveguides. As no experimental evidence of Mg clustering 
in LN is available, we assume a homogeneous distribution. 

The position of Li$^+$ and Mg$^{2+}$ within the oxygen octahedra is different.
Although the distance to the three oxygen ions below them is roughly the same
(2.013\,{\AA} and 2.022\,{\AA} for Li$^+$ and Mg$^{2+}$, respectively), the distance
to the three oxygen ions above them is much shorter for Mg$^{2+}$ (2.172\,{\AA}) than
for Li$^+$ (2.291\,{\AA}). Thus, the oxygen octahedra occupied by the
Mg$^{2+}_{\mathrm{Li}}$ substitutional is more contracted than
the octahedron occupied by a regular Li$^+$ ion. The contraction is also visible
in the oxygen-oxygen interatomic distance, which is much shorter in the Mg$^{2+}$
octahedron (3.298 {\AA} vs 3.356 {\AA}), as shown in Figure \ref{fig:mg_geom}.
In this configuration,
Mg doping does not induce localized defect states in the LiNbO$_3$ bandgap. The 
calculated band structure of Mg doped LiNbO$_3$ is shown in the SI.

\begin{figure}
  \includegraphics[width=0.46\linewidth]{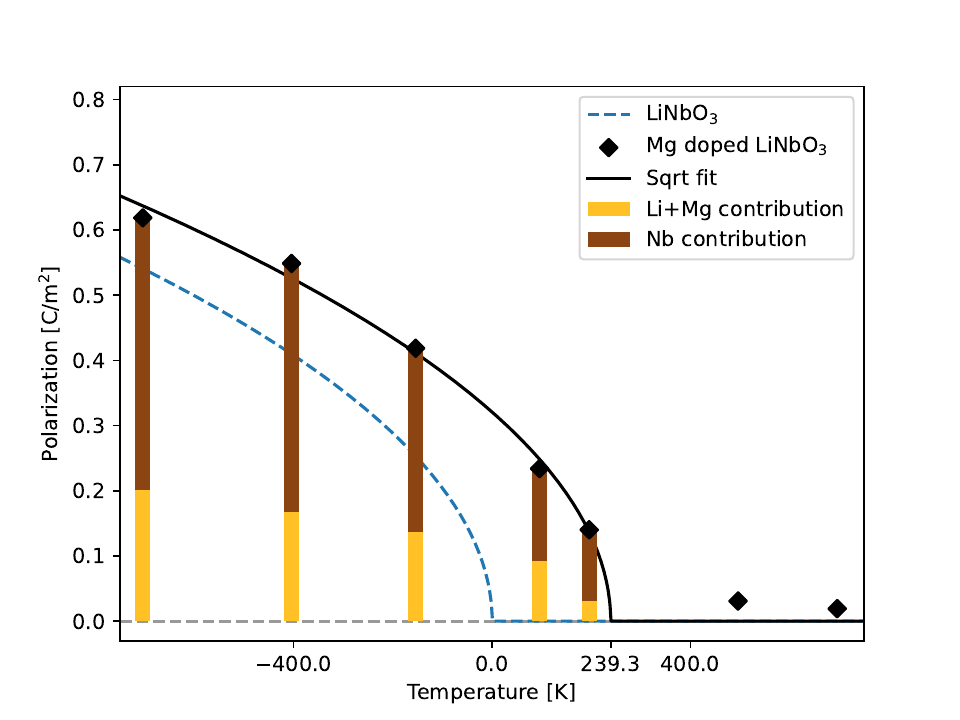}
  \caption{Calculated temperature dependence of the spontaneous polarization $P_S$ for 
    Mg doped (solid line) and undoped (dashed line) LiNbO$_3$. The solid line is a square root fit of 
    the calculated data, represented by the histograms. The temperature is given with respect to the
    calculated transition temperature of undoped LiNbO$_3$. \label{fig:polarisationMg}}
\end{figure}

The AIMD runs of supercells containing Mg show the same qualitative trend as the
supercells modelling the undoped material. The probability distributions of $\Delta$Li 
and $\Delta$Nb have a similar form and temperature dependence as in the case of the undoped 
crystals, however, they are shifted to higher temperatures. Correspondingly, 
the spontaneous polarization vanishes at temperatures of about 240\,K higher,
as shown in Figure \ref{fig:polarisationMg}. This value, which must be considered as 
a rough estimate, is in qualitative agreement with the 
experimental observation. For a doping 
of 1.2 atom \%, a Curie temperature of 1493\,$^\circ$C is measured \cite{Furukawa01}, 
which is 80\,K higher than the value of 1413\,$^\circ$C measured for the undoped samples
\cite{Chen01}. 

\begin{figure}
  \includegraphics[width=0.8\textwidth]{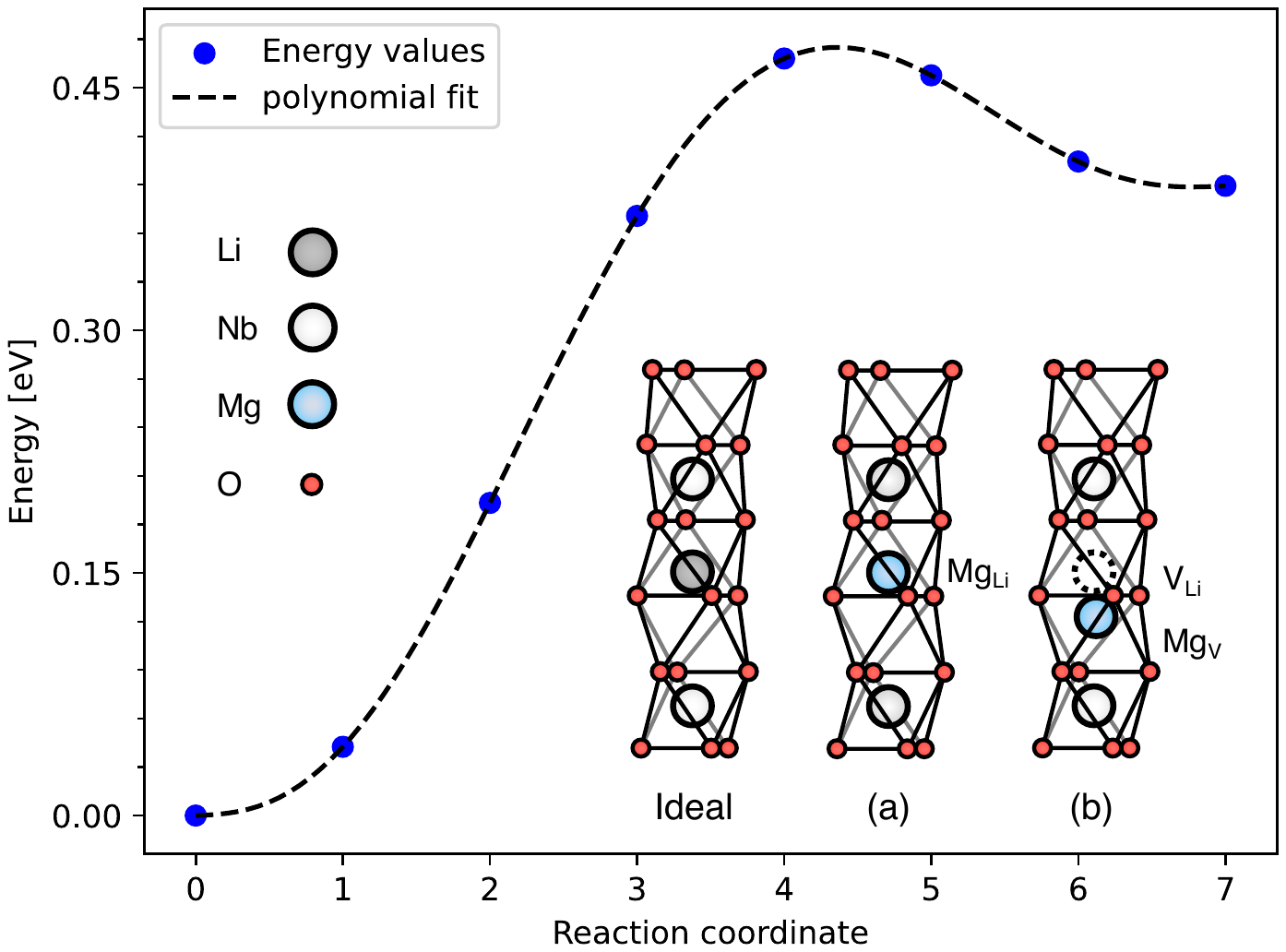}
  \caption{Energy barrier associated with the Mg$^{2+}_{\text{Li}}$ diffusion to an empty octahedron,
    calculated for LiNbO$_3$ with the NEB method and six images besides start and end configuration.
    Dotted lines are polynomial fit of the calculated data. The inset shows the undoped crystal structure
    as well as start and end configuration of the NEB.}\label{fig:Migration}
\end{figure}

This effect can be explained by different factors. The Mg ions pin the polarization 
locally, as they require more energy than Li ions to pass the oxygen layer. 
The minimum energy path for the migration of the Mg$^{2+}_{\mathrm{Li}}$ substitutional
to the empty octahedron interstitial position has been calculated by 
the NEB method, considering 6 images between start and end configuration. 
Start and end configuration are shown in the inset of Figure \ref{fig:Migration}.

The calculated energy barrier of 0.47\,eV is  
roughly an order of magnitude higher than the barrier that we calculate, e.g., for a similar 
migration path of Li in hydrogenated LiNbO$_3$. 
The reason for such a striking difference of the barrier heights
cannot be merely geometric, as Mg and Li have nearly identical ionic radii (0.90\,{\AA} for 
Li$^+$ and 0.86\,{\AA} for Mg$^{2+}$ in octahedral coordination \cite{webelements}). 
However, the position of the two ions within the oxygen octahedra is different. 
In particular, the oxygen octahedra occupied by the  Mg$^{2+}_{\mathrm{Li}}$ substitutional
is more contracted than the octahedron occupied by a regular Li$^+$ ion, probably due to 
the larger Coulomb attraction between the cation and the anions.
Thus, the oxygen ions in the basal plane that has to be crossed during the phase transion
build a tighter mesh for Mg$^{2+}$ (O-O distance 3.298\,{\AA}) than for Li$^{2+}$ 
(O-O distance 3.356\,{\AA}), as shown in Figure \ref{fig:mg_geom}. 
We thus suggest that on the one hand a larger deformation of the
oxygen octahedra, and on the other hand a stronger Mg-O bond than the Li-O bond,
both contribute to locally pin the polarization and thus the ferroelectric phase.

\subsection{\label{sec:res_exp}Comparison with experimental results} 

Although the calculated Curie temperatures reasonably match the experimental data,  
it becomes clear from the
AIMD that the atomic positions start to shift at temperatures well below $T_C$.
An interval of about 100\,K and 300\,K for LiTaO$_3$ and LiNbO$_3$, respectively,
can be identified, in which the largest part of the structural modifications occur. 

In order to define more closely this interval, we compare the calculated value of 
the spontaneous polarization as well as the octahedra occupation with 
physical quantities that can be observed experimentally by characterizing the crystals
across the phase transition. 
In the following, we start our discussion with LiTaO$_3$, which is easier to 
characterize experimentally due to the lower $T_C$.

\begin{figure}[t]
\includegraphics[width=0.6\linewidth]{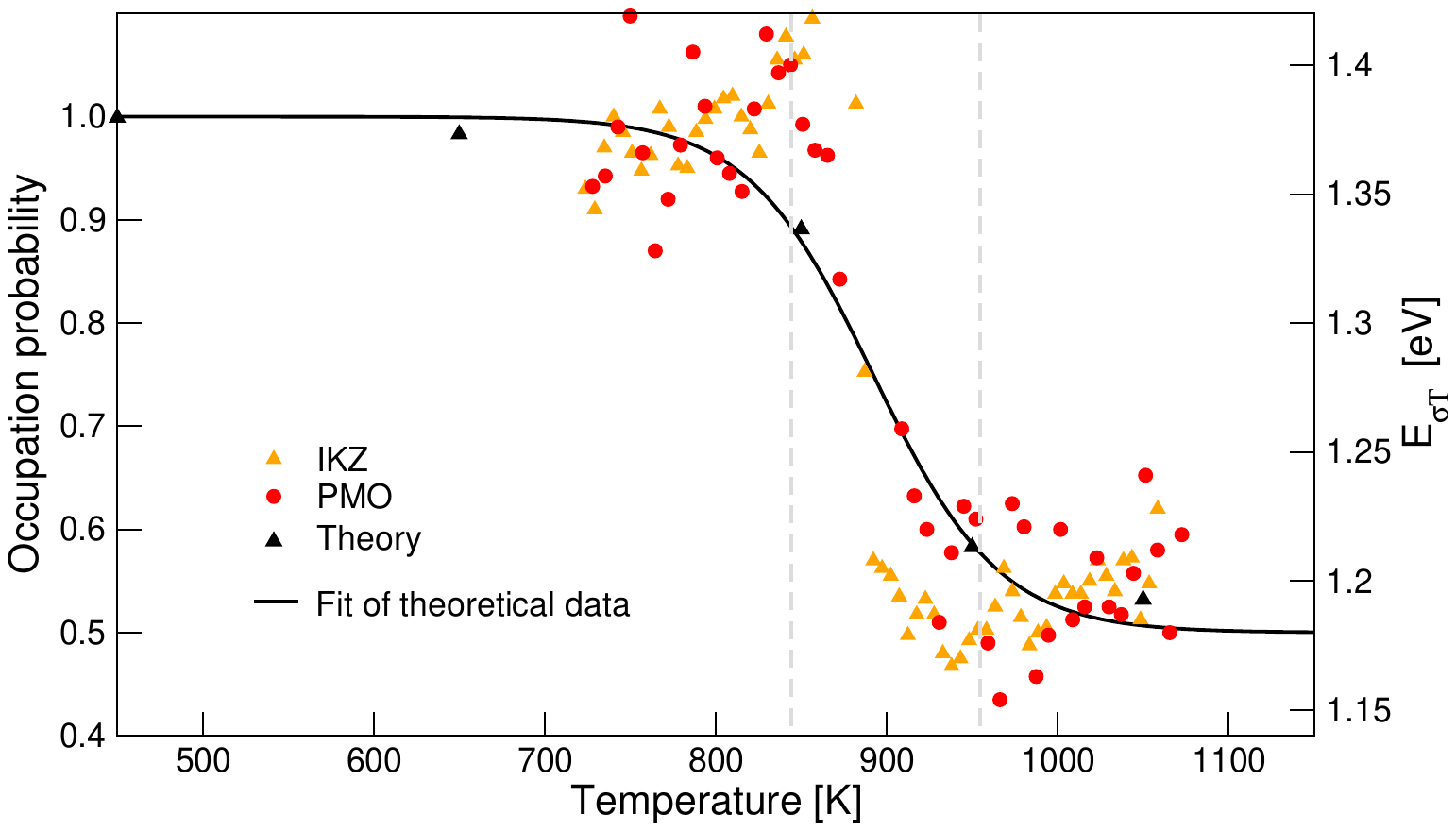}
\caption{Calculated temperature dependent occupation of the regular Li octahedra 
(black triangles), fit through a sigmoid function (solid line) and measured 
slope of the electrical conductivity of LiTaO$_3$.
Note that the scatter of $E_{\sigma T}$ is caused by the calculation 
of the slope of the electrical conductivity using equation \ref{eq:slope}.
For the sake of clarity, the calculated data is rigidly translated to match the measured $T_C$.
The samples used for the measurements are named according to the description in the conductivity 
section.\label{fig:conductivity}}
\end{figure}

To characterize the phase transition, we define the occupation of the regular Li octahedra 
as a reaction coordinate. 
In the ferroelectric phase and at very low temperatures, the occupation of the regular Li 
octahedra is roughly 100\%. For increasing temperature, the Li atoms may possess enough thermal energy 
to overcome the energy barrier represented by the oxygen plane and migrate into the vacant oxygen
octahedra (see, e.g., figure \ref{fig:crystals}). In the paraelectric phase, the Li atoms jump
continuously between the Li octahedra and the vacant octahedra, resulting in an occupation of 50\% of the 
regular Li octahedra. The occupation of the regular Li octahedra is shown in figure 
\ref{fig:conductivity} (black triangles). The solid line is a fit of the calculated data
by a sigmoid function (fits with other functions are possible as well) which only serves 
as a guide to the eye.

The interval at which the occupation of the Li octahedra changes from 100\%
to 50\% is an estimate of the temperature interval at which the phase transition occurs.
In the case of LT, the width of this interval is roughly 100\,K.

It is known that the electrical conductivity of LT changes 
at the transition from the ferroelectric to the paraelectric phase, at which
the measured activation energy decreases from a value of 1.38\,eV to 1.19\,eV \cite{UlianaDez23}. 
Accurate experiments 
have been performed on different PMO samples (see experimental details) 
to explore the temperature range around $T_C$ and establish whether the octahedra occupation
correlates with the crystal conductivity. The corresponding 
measurements are shown in figure \ref{fig:conductivity}. 
The temperature interval at which the slope $E_{\sigma T}$ varies from 1.38\,eV to 1.19\,eV roughly corresponds 
to the theoretically predicted range, confirming that the phase transition is a continuous 
process extending over a larger temperature interval. The comparison between calculated and measured data also
shows a correlation between octahedra occupation and conductivity. The analysis of this
correlation is, however, beyond the scope of this manuscript and is discussed elsewhere \cite{FatimaDez23}.
Although the measurement clearly show that the conductivity starts to change well below the Curie temperature, 
the scattering of the measured data (due to the calculation of the derivative of the conductivity) 
does not allow a more precise estimate of the interval width at which the transformation occurs.

\begin{figure}[t]
\includegraphics[width=0.6\linewidth]{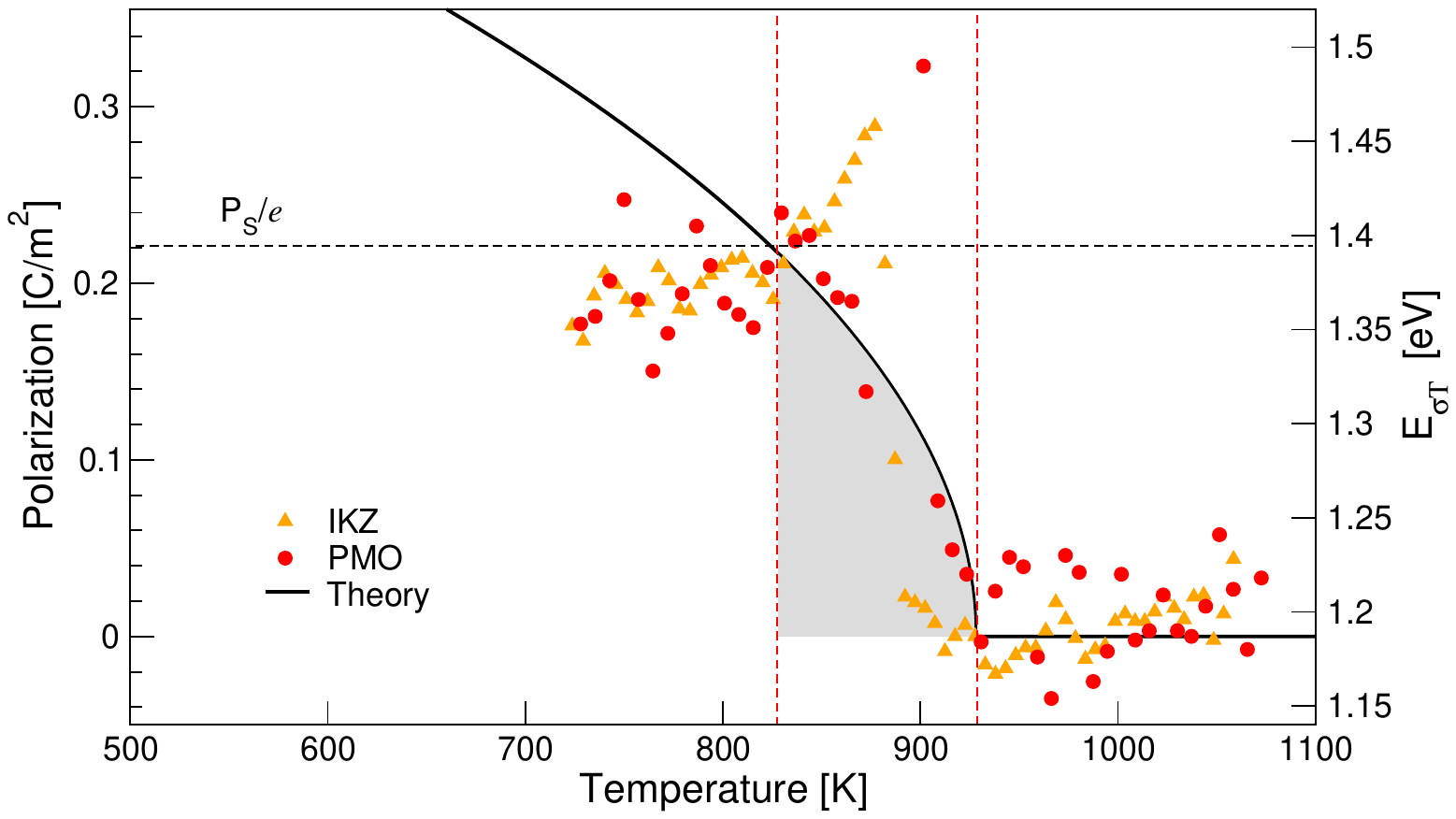}
\caption{Calculated spontaneous polarization as a function of the temperature  (solid line) 
and measured slope of the electrical conductivity of LiTaO$_3$. For the sake of clarity, the calculated 
data is rigidly translated to match the measured $T_C$. The samples used for the measurements 
are named according to the conductivity section. The gray region marks the temperature interval
at which the polarization has decreased below a factor $1/e$ of the 0\,K value.
\label{fig:conductivity_LT}}
\end{figure}

For that reason, we correlate the measured conductivity with the spontaneous polarization as a
function of the temperature, as shown in figure \ref{fig:conductivity_LT}. Indeed, the order 
parameter has at least a well defined point at which it vanishes, namely $T_C$, which can be 
considered the end of the temperature interval at which the transition occurs. The starting
point can be arbitrarily chosen to be the point at which the order parameter has grown to 
a factor $1/e$ of the 0\,K value. This region is marked in gray in figure \ref{fig:conductivity_LT}
and corresponds to about 100\,K.
Interestingly, the conductivity jumps roughly in this temperature range from the value
of the paraelectric to the value of the ferroelectric phase. This confirms, again, that the
geometry modification starts to affect the material properties below $T_C$. 

\begin{figure}[t]
\includegraphics[width=0.6\linewidth]{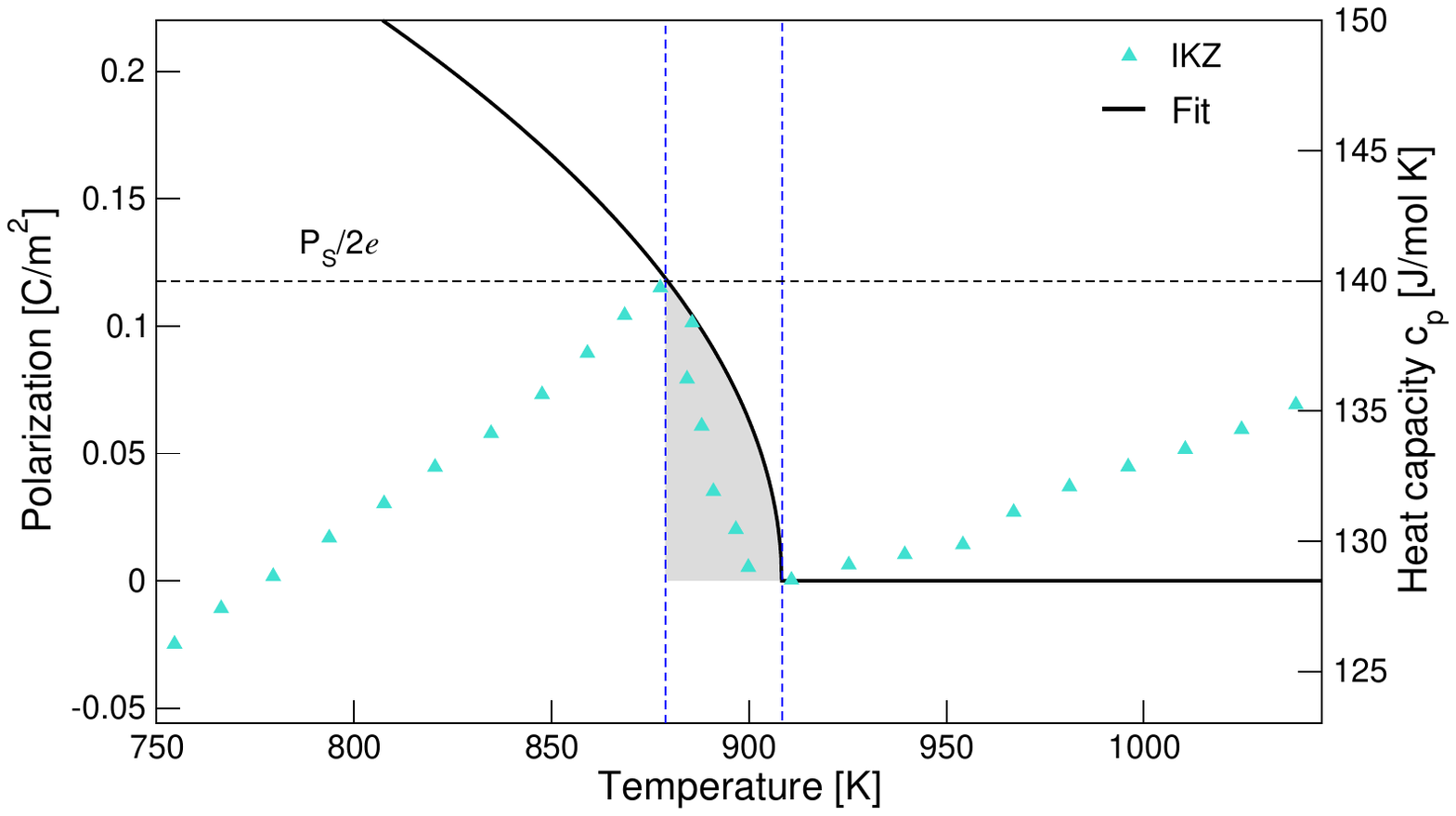}
\caption{Calculated temperature dependence of the spontaneous polarization (solid line) 
and measured thermal capacity at constant pressure $C_p$ of LiTaO$_3$ (light blue triangles). 
For the sake of clarity, the calculated data is rigidly translated to match the measured $T_C$.
The gray region marks the temperature interval                                                              
at which the polarisation has decreased below a factor $1/2e$ of the 0\,K value.
\label{fig:capacity_LT}}
\end{figure}

Another thermodynamical quantity which is expected to be affected by the phase transition 
is the heat capacity. As the ferroelectric and the paraelectric phase have different 
symmetry, also the phonon population is different and with it the capacity of the material 
to store energy. In order to verify in which temperature range the thermal capacity of 
LiTaO$_3$ jumps from the value measured for the ferroelectric phase to the value measured 
for the paraelectric phase, calorimetry
experiments have been performed. The corresponding measurements are shown in figure
\ref{fig:capacity_LT} and again compared with the theoretically predicted order parameter.
The jump of $C_p$ takes place in a temperature range which is narrower than the range at which 
the octahedra occupation or the conductivity changes. The involved temperature range corresponds
to the interval at which the spontaneous polarization grows from zero to a value roughly $1/2e$ 
smaller than the 0\,K value. Thus, although the measurements corroborate the hypothesis 
that the structural transition is a continuous process affecting the materials properties already
below $T_C$, the temperature interval depends on how the microscopic structure affects the 
measured quantity.

\begin{figure}[t]
\includegraphics[width=0.4\textwidth,trim=25 100 215 80,clip]{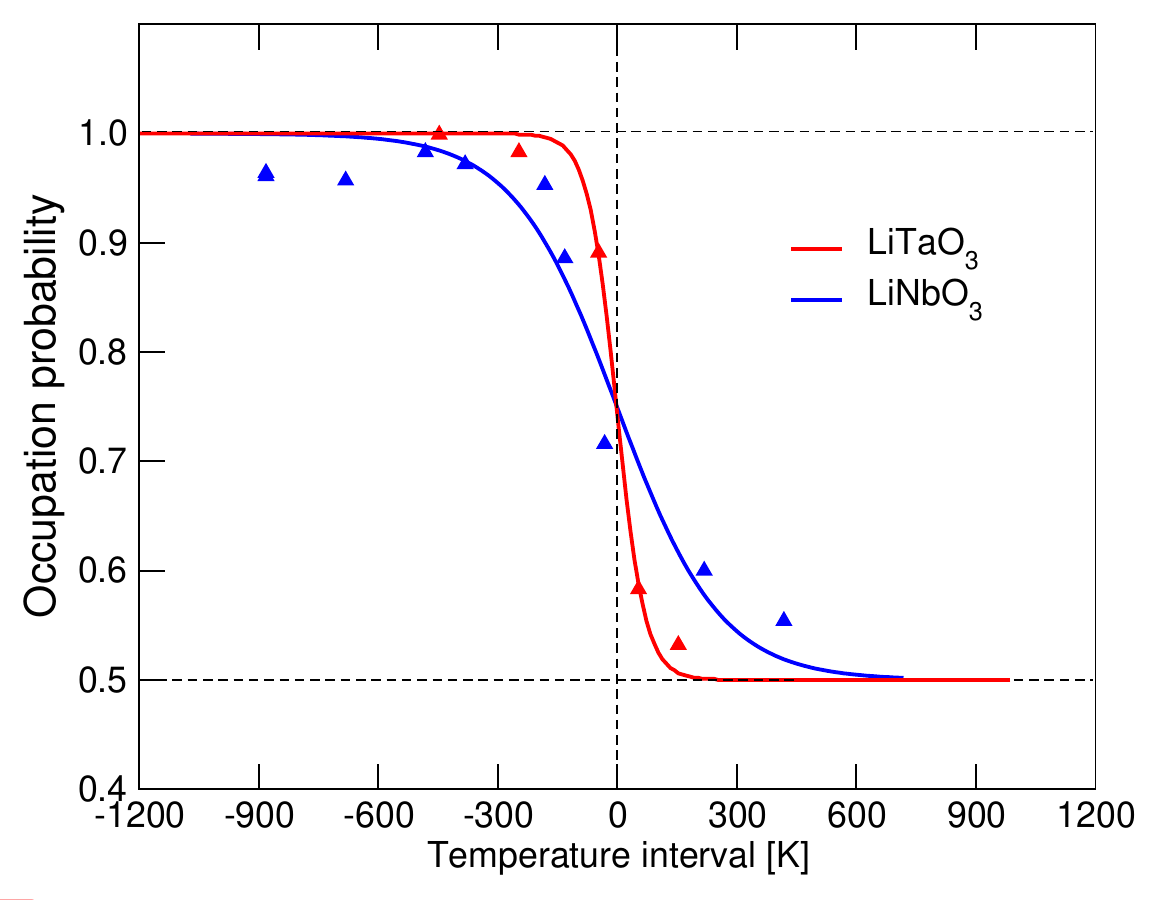}
\caption{Comparison of the calculated temperature dependence of the occupation of the regular
Li octahedra in LiTaO$_3$ (red triangles) and LiNbO$_3$ (blue triangles). The solid line is a 
sigmoid function fit of the calculated data. The data is shifted, so that the inflection point
of each curve is at the origin of the temperature axis.\label{fig:comparison}}
\end{figure}

A comparison of the experimental data from transport and calorimetry measurements 
with the theoretically predicted temperature dependent order parameter (e.g., the spontaneous
polarization) can be found in the SI. Also in the case of LiNbO$_3$ it becomes evident 
that similarly to LiTaO$_3$ the atomic displacements have an influence on the conductivity
and on the heat capacity below $T_C$. Again, the temperature interval depends on the 
investigated property. 
Finally, the octahedra occupation of LiTaO$_3$ and LiNbO$_3$ as a function of temperature 
is compared. It might be supposed that the energy needed from the Li atoms to pass the oxygen
plane is of the same order of magnitude in the two compounds. It is therefore 
reasonable to assume that 
the Li ions begin their migration at similar temperatures in the two crystals. However, as the
Curie temperature of LN is much higher than the Curie temperature of LT, it is expected that
the structural transition in LN occurs in a somewhat larger temperature interval. The AIMD
trajectories confirm this hypothesis. Figure \ref{fig:comparison} shows the occupation of
the Li octahedra for the two compounds. While in LT the occupation sinks from 100\% to 50\%
within 100\,K, in LN the same occurs within a substantially larger temperature range of 
about 300\,K. 

Finally, we remark that, besides doping, also the intrinsic defect structure 
will affect the ferroelectric
phase transition in LiNbO$_3$ and LiTaO$_3$. In LiNbO$_3$, specific structural defects are known, that 
are likely to be formed \cite{YanluOptic,Xu08}. In particular, a Nb$_{\mathrm{Li}}$ antisite 
that is charge compensated by four Li vacancies (Li vacancy model) is currently assumed to be 
the dominant defect cluster. In LiTaO$_3$, the defect structure is more complex, yet a Ta 
at an interstitial position and charge compensated by five Li vacancies has been identified 
as an energetic favorable structure \cite{Vyalikh08}. The dipole moments associated with the 
defect clusters will locally affect the materials properties and locally pin the polarization. 
The diffusivity of the lithium vacancies is low at room temperature, which indicates that the 
defect complexes are rather stable. Even at high temperatures, although the Li vacancies become 
very mobile \cite{Claudia23}, the reported cohesive energy of above 4\,eV of the clusters 
\cite{YanluOptic} suggests that the clusters may still affect the transition temperature by 
locally pinning the crystal structure. The local pinning will result in an extension of the 
temperature range of the transition temperature. A quantitative analysis of this effect is 
beyond the goals of this investigation.

\section{\label{sec:summary}Summary and outlook}

The ferroelectric to paraelectric phase transition in LiTaO$_3$ and LiNbO$_3$ has been investigated 
theoretically and experimentally. First principle calculations in the framework of the SSCHA 
formalism predict a transition temperature of 808\,K and 1408\,K, for LiTaO$_3$ and LiNbO$_3$,
respectively, which are in rather good agreement with the measured values.
AIMD calculations give insight into the mechansims of the phase transition.
The latter is a complex process, in which Li and Nb/Ta behave differently. 
The Li sublattice undergoes an order-disorder structural transition, while the Nb/Ta-sublattice undergoes a 
displacive type transition. For both LiTaO$_3$ and LiNbO$_3$, substantial atomic displacements 
are predicted for temperatures smaller than $T_C$. From an atomistic perspective, 
we interpret the transition as a structural modification, whose magnitude is a function of the 
temperature and which results in a non-vanishing polarization at $T_C$. 
Starting from the transition temperature and moving towards lower temperatures, 
the polarization continuously grows. On top of this structural modification, the presence of structural defects and 
defect clusters will affect the atomic movement.

The structural modifications are detectable, to a certain extent,
in transport and calorimetry experiments. Experimental measurements of the conductivity 
of both LT and LN demonstrate signatures of the structural modifications well below $T_C$, 
while the heat capacity is modified in a narrower interval, instead.

Finally, the Curie temperature of LN is found to be raised by Mg doping.
Mg$_{\mathrm{Li}}$ substitutionals are found to locally pin the ferroelectric
polarization, thus acting against thermal disorder.
The presented AIMD calculations allow to investigate how dopants influence the 
Curie temperature of the samples and can be applied to other common dopants such 
as Ti, Fe, or Ni. 

Another field in which a similar approach is expected to be of large benefit concerns
the study of LNT solid solutions. The solid solutions, grown specifically to combine
the optical and acoustic properties of LN with the thermal stability of LT, represent
a rather new field of research, for which the characterization of the thermal behavior
will be of crucial relevance.



\begin{acknowledgments}
We gratefully acknowledge financial support by the Deutsche Forschungsgemeinschaft (DFG) through
the research group FOR5044 (Grant No. 426703838 \cite{FOR5044}, SA1948/3-1, SU1261/1-1, SCHM 1569/39-1, FR1301/42-1). 
Calculations for this research were
conducted on the Lichtenberg high-performance computer of the TU Darmstadt and at the 
H\"ochstleistungrechenzentrum Stuttgart (HLRS). The authors furthermore acknowledge the
computational resources provided by the HPC Core Facility and the HRZ of the Justus-Liebig-Universit\"at 
Gie{\ss}en.
\end{acknowledgments}

\bibliography{literature_LNT}

\end{document}